\documentclass[12pt,epsfig]{article} 
\pdfoutput=1
\usepackage{amssymb, amsmath,mathrsfs}
\usepackage{graphicx}
\usepackage{subfig}
\usepackage{multirow}
\usepackage{cite}
\usepackage{chngcntr}
\usepackage[colorlinks=true,
            linkcolor=red,
            urlcolor=blue,
            citecolor=blue]{hyperref}
\usepackage{dcolumn}
\usepackage{bm}
\usepackage{color}
\long\def\rpl#1!!#2!!{\textcolor{red}{#1} \textcolor{blue}{#2}}
\def\baselinestretch{1.3}
\newcommand{\ba}{\begin{array}}
\newcommand{\ea}{\end{array}}
\newcommand{\bd}{\begin{displaymath}}
\newcommand{\ed}{\end{displaymath}}
\newcommand{\besub}{\begin{subequations}}
\newcommand{\eesub}{\end{subequations}}
\newcommand{\be}{\begin{equation}}
\newcommand{\ee}{\end{equation}}
\newcommand{\bea}{\begin{eqnarray}}
\newcommand{\eea}{\end{eqnarray}}
\newcommand{\no}{\nonumber\\}

\def\a{\alpha}

\def\b{\beta}

\def\l{\lambda}

\def\L{\Lambda}
\def\s{\sigma}

\def\q2 {q^2}

\def\bt{\begin{table}}
\def\et{\end{table}}

\catcode`@=11 
\def \gsim{\mathrel{\mathpalette\@versim>}}
\def \lsim{\mathrel{\mathpalette\@versim<}}
\def \@versim#1#2{\lower0.4ex\vbox{\baselineskip\z@skip\lineskip\z@skip
     \lineskiplimit\z@\ialign{$\m@th#1\hfil##\hfil$
     \crcr#2\crcr\sim\crcr}}}
\catcode`@=12 

\allowdisplaybreaks
\begin{document}

\begin{flushright}
{HRI-RECAPP-2014-016}
\end{flushright}

\begin{center}

{\large \textbf {High-scale validity of a
two-Higgs doublet scenario: a study including LHC data}}\\[15mm]

Nabarun Chakrabarty$^{\dagger}$\footnote{nabarunc@hri.res.in}, 
Ujjal Kumar Dey$^{\dagger}$\footnote{ujjaldey@hri.res.in}, and  
Biswarup Mukhopadhyaya$^{\dagger}$\footnote{biswarup@hri.res.in}\\
$^{\dagger}${\em Regional Centre for Accelerator-based Particle Physics \\
     Harish-Chandra Research Institute\\
 Chhatnag Road, Jhunsi, Allahabad - 211 019, India}\\[5mm] 

\end{center}

\begin{abstract} 

 We consider the conditions for the validity of a two-Higgs doublet model
at high energy scales, together with all other low- and high-energy
constraints. The constraints on the parameter space at low energy,
including the measured value of the  Higgs mass and the signal strengths
in channels are juxtaposed with the conditions of vacuum stability,
perturbativity and unitarity at various scales. We find that a scenario with
an exact $\mathbb{Z}_2$ symmetry in the potential cannot be valid beyond about
10 TeV without the intervention of additional physics. On the other hand,
when the $\mathbb{Z}_2$ symmetry is broken, the theory can be valid even up to the
Planck scale without any new physics coming in. The interesting feature we
point out is that such high-scale validity is irrespective of the uncertainty in the top
quark mass as well as $\a_s(M_Z)$, in contrast with the standard model with a single Higgs doublet. It is also shown that the presence of a CP-violating phase is
allowed when the $\mathbb{Z}_2$ symmetry is relaxed. The allowed regions in the
parameter space are presented for each case. The results are illustrated in the context of a Type-II scenario.

\end{abstract}

\newpage
\setcounter{footnote}{0}

\def\baselinestretch{1.5}
\counterwithin{equation}{section}
\section{Introduction}

The Higgs sector of the standard electroweak model (SM) continues to
appear enigmatic from several angles. The existence of such a sector, comprising at least one scalar doublet, and driving the spontaneous symmetry breakdown $SU(2)_L \times U(1)_Y \longrightarrow U(1)_{EM}$ is almost impossible to deny now. It is also widely agreed that the Large Hadron Collider (LHC) has found \cite{Aad:2012tfa,Chatrchyan:2012ufa} a neutral boson with
mass around 125 GeV, which is almost certainly of spin zero~\cite{Aad:2013xqa} and
dominantly a CP-even field~\cite{ATLAS:2013nma,Freitas:2012kw, Djouadi:2013yb, Djouadi:2013qya}. However, despite the properties of the
boson being {\em consistent} with that of the SM Higgs, rather
persistent enquiries are on, to find out whether the electroweak
symmetry breaking sector also contains some signature of physics
beyond the standard model. The LHC data till date leaves room for such
new physics.

Two sets of standpoints are noticed in such enquiries.  First of all,
with spin-1/2 fermions showing family replication, it is not obvious
why the part of the matter sector containing spin-zero particles
should also not have similar repetition. With this in view,
multi-doublet scenarios are under regular scrutiny, the most widely
investigated models being those with two Higgs doublets\cite{Gunion:1989we,Branco:2011iw}. An extended electroweak symmetry breaking sector entails a rich phenomenology, including additional sources of CP violation~\cite{Deshpande:1976yp}. Of course, scalars
belonging to higher representations of $SU(2)$ have also attracted
attention, especially triplets which can play a role in the so-called
Type-II mechanism of neutrino mass generation~\cite{Mohapatra:1979ia}.  Secondly, even with
just one doublet (leading to a single physical scalar), the Higgs mass
is not stable under quadratically divergent radiative corrections, and
it is somewhat artificial (or `fine-tuned') to have a 125 GeV Higgs
if the cut-off for the SM is much higher than a TeV or so.
Furthermore, it is also not clear that the SM scalar potential retains
a finite and stable minimum at high scales. But for the yet uncertain
measurement of the top quark mass, which is crucial in governing the
evolution of the Higgs self-coupling via Yukawa interactions, we may
be doomed to live in an unstable or metastable vacuum if no new physics intervenes within the scale $10^{8-10}$ GeV~\cite{Holthausen:2011aa, EliasMiro:2011aa, Degrassi:2012ry, Zoller:2012cv,Buttazzo:2013uya}. Therefore, the ultraviolet
incompleteness of the current scheme of electroweak symmetry breaking
looms up as a distinct possibility, even if one disregards the somewhat 
philosophical issue of naturalness.

In this paper, we follow these two standpoints in tandem.  We take up
a two-Higgs doublet scenario as the minimal extension of the standard
electroweak theory, assessing its viability as well as sufficiency modulo
all available constraints. The motivation for the study is 
that the proportionality constant between the top quark mass and its
coupling to the 125 GeV scalar is different from its SM value when
more than one doublet is taken. Consequently, the dependence of the vacuum
stability limit on the top quark mass is expected to be different.
However, one can make precise and quantitative statements on the matter
only when one takes cognizance of the exact scenario, and includes
the complete set of renormalisation group equations appropriate 
for it. This is precisely what we aim to do here, using a two-Higgs doublet
scenario at various levels of generality.

The desired suppression of flavour-changing
Yukawa interactions is best implemented by imposing a discrete
symmetry on such models, thus preventing both the doublets from
coupling with $T_3 = +1/2$ and $-1/2$ fermions simultaneously. It is
possible to go beyond such imposition and examine two Higgs doublets
in a `basis-independent' formulation~\cite{Davidson:2005cw, Haber:2006ue, Haber:2010bw}.  However, we feel that our
central issue, namely, the evolution of the Higgs self-interaction(s),
is amenable to a more transparent study if one adheres to a specific
Yukawa scheme. With this in view, we adopt the so-called Type-II
scenario for our study~\cite{Gunion:1989we}, to illustrate our point.

We begin by examining the situation when the discrete symmetry is
exact, and derive the constraints on the low-energy values of the
parameters of this scenario. The lighter neutral scalar mass being
around 125 GeV is of course the prime requirement here, and
constraints from rare processes such as $b \rightarrow s \gamma$
are also included. In addition, the constraints from 
perturbativity of all scalar quartic couplings are considered,
together with those from vacuum stability.  The parameter space thus
validated is further examined in the light of the perturbativity and
vacuum stability conditions at high scales. Thus we identify the
parameter regions that keep a two-Higgs doublet scenario valid up to
different levels of high scales-- an exercise that reveals rather
severe limits.  The same investigation is carried out for cases where
the discrete symmetry is broken by soft (dimension-2) and hard
(dimension-4) terms in turn, with the Yukawa coupling assignment
remaining (for simplicity) the same as in the case with unbroken
symmetry. The effect of a CP-violating phase is also demonstrated.
Finally, the regions found to be allowed from all the above
considerations, at both low- and high-scales, are pitted
against the existing data from the LHC in different channels. Thus we
identify parameter regions that are consistent with the measured
signal strengths in different channels. This
entire study is aimed at indicating how far a two-Higgs doublet model
can remain valid, not only at the LHC energy but also up to various high
scales without further intervention of new physics.

Although a number of recent studies have addressed some similar questions\cite{Barroso:2013awa,Gorczyca:2011he,Kannike:2012pe},
the present study has gone beyond them on the following points:

\begin{itemize}
\item Our study reveals that the high scale validity of the theory is less sensitive to the precise value of the top quark mass than in the SM. Regions in the parameter space are identified, for which the theory has no cut-off till the Planck scale, even though the top quark mass can be at the upper edge of the allowed band. Similarly, the high scale validity of the model is insensitive to $\alpha_{s}(M_{Z})$.

    \item We find that it is rather difficult to retain the validity
      of a two-Higgs doublet scenario well above a TeV with the
      discrete symmetry intact. Also, large values of $\tan\beta$,
      the ratio of the vacuum expectation values (vev) of the two
      doublets, are mostly disfavoured in this case.

    \item With the discrete symmetry broken, the theory can circumvent ultraviolet cut-offs. There is a correlation between allowed $\tan\beta$ and the extent of symmetry breaking, when it comes to validity up to the Planck scale.

    \item We examine the constraints on the model including a
          CP-violating phase\cite{Kopp:2014rva,Inoue:2014nva,Arhrib:2010ju,Osland:2013sla}. In fact, since the   existence of a phase is a
          natural consequence of relaxing the discrete symmetry, the
          high-scale validity of a two Higgs doublet model may be argued
          to be contingent on the possibility of CP-violation in the
          scalar potential.
       
        \item We have performed a detailed examination of the validity
          of the scenario at both low- and high scales, {\em including
            dimension-4 discrete symmetry breaking terms
            in our analysis.} The LHC constraints are also imposed in
          this situation.
\end{itemize}

We remind the reader of the broad features of a two-Higgs doublet
scenario in section 2.  In section 3, we list and explain all the
constraints that the scenario is subjected to, at both the low and
high scales. Sections 4, 5 and 6 contain, in turn, the results of our
analysis, with the discrete symmetry intact, softly broken and broken
by hard terms, respectively.  We summarise and conclude in section 7.

\section{The two-Higgs-doublet scenario and the scalar potential:
basic features}

In the present work, we consider the most general renormalisable
scalar potential for two doublets $\Phi_1$ and $\Phi_2$, each having
hypercharge $(+1)$,
\bea
V(\Phi_1,\Phi_2) &=&
m^2_{11}\, \Phi_1^\dagger \Phi_1
+ m^2_{22}\, \Phi_2^\dagger \Phi_2 -
 m^2_{12}\, \left(\Phi_1^\dagger \Phi_2 + \Phi_2^\dagger \Phi_1\right)
+ \frac{\lambda_1}{2} \left( \Phi_1^\dagger \Phi_1 \right)^2
+ \frac{\lambda_2}{2} \left( \Phi_2^\dagger \Phi_2 \right)^2
\no & &
+ \lambda_3\, \Phi_1^\dagger \Phi_1\, \Phi_2^\dagger \Phi_2
+ \lambda_4\, \Phi_1^\dagger \Phi_2\, \Phi_2^\dagger \Phi_1
+ \frac{\lambda_5}{2} \left[
\left( \Phi_1^\dagger\Phi_2 \right)^2
+ \left( \Phi_2^\dagger\Phi_1 \right)^2 \right]
\no & &
+\lambda_6\, \Phi_1^\dagger \Phi_1\, \left(\Phi_1^\dagger\Phi_2 + \Phi_2^\dagger\Phi_1\right) + \lambda_7\, \Phi_2^\dagger \Phi_2\, \left(\Phi_1^\dagger\Phi_2 + \Phi_2^\dagger\Phi_1\right).
\label{pot}
\eea

The parameters $m_{12}$, $\lambda_5$, $\lambda_{6}$ and $\lambda_7$
could be complex in general, although the phase in {\em one} of them
can be removed by redefinition of of the relative phase between
$\Phi_1$ and $\Phi_2$.  Thus this scenario in general has the
possibility of CP-violation in the scalar sector.

In a general two-Higgs-doublet model (2HDM), a particular fermion can
couple to both $\Phi_1$ and $\Phi_2$. However this would lead to the
flavor changing neutral currents (FCNC) 
at the tree level\cite{PhysRevD.15.1958,PhysRevD.15.1966,Arhrib:2005nx,Dery:2013aba}\footnote{In context of a typical flavour changing scenario, it has been shown in \cite{Cvetic:1997zd,Cvetic:1998uw} that the FCNCs are stable under RG evolution to a fairly large degree.}.
One way to avoid such FCNC is to impose a $\mathbb{Z}_2$ symmetry, such as one
that demands invariance under $\Phi_1 \to -\Phi_1$ and $\Phi_2 \to
\Phi_2$. This type of symmetry puts restrictions on
the scalar potential. The $\mathbb{Z}_2$ symmetry is \emph{exact} as
long as $m_{12}$, $\lambda_{6}$ and $\lambda_7$ vanish, when the scalar
sector also becomes CP-conserving. The symmetry
is said to be broken \emph{softly} if it is violated in the quadratic
terms only, i.e., in the limit where $\lambda_{6}$ and $\lambda_7$
vanish but $m_{12}$ does not. Finally, a \emph{hard} breaking of the
$\mathbb{Z}_2$ symmetry is realized when it is broken by the quartic
terms as well. Thus in this case, $m_{12}$, $\lambda_{6}$ and
$\lambda_7$ all are non-vanishing in general.
 
As mentioned in the introduction, we focus on a specific scheme
of coupling fermions to the doublets. This scheme is referred to in
the literature as the \emph{Type-II} 2HDM, where the down type quarks
and the charged leptons couple to $\Phi_1$ and the up type quarks, to
$\Phi_2$\cite{Pich:2009sp}. This can be ensured through the discrete symmetry $\Phi_{1}\to -\Phi_{1}$ and $\psi_{R}^{i}\to -\psi_{R}^{i}$, where $\psi$ is charged leptons or down type quarks and $i$ represents the generation index.  Although we start by analysing the high-scale validity 
of the model with $m_{12} = \lambda_{6} = \lambda_7 = 0$, we subsequently 
include the effects of both soft and hard breaking of $\mathbb{Z}_2$
in turn, which bring back these parameters. The two simplifications
that we still make are as follows: (a) the phases of $\lambda_{6}$ and
$\lambda_7$ are neglected though that of $m_{12}$ is considered, and
(b) the Yukawa coupling assignments of $\Phi_1$ and $\Phi_2$ are 
left unchanged.  

Minimization of the scalar potential in Eq.~\ref{pot} yields 
\be
\langle \Phi_1 \rangle
= \left( \begin{array}{c} 0 \\ \displaystyle{\frac{v_1}{\sqrt{2}}} \end{array} \right),
\quad
\langle \Phi_2 \rangle
= \left( \begin{array}{c} 0 \\ \displaystyle{\frac{v_2}{\sqrt{2}}} \end{array} \right),
\ee  
\noindent
where the vacuum expectation values (vev) are often expressed in terms of 
the $M_Z$ and the ratio
\be
\tan \beta = \frac{v_2}{v_1}\;. 
\ee
We parametrise the doublets in the following fashion,
\be
\Phi_{i} = \frac{1}{\sqrt{2}} \begin{pmatrix}
\sqrt{2} w_i^{+} \\
v_i + h_i + i z_i
\end{pmatrix}~ \rm{for}~\textit{i} = 1, 2.
\label{e:doublet}
\ee
Since the basis used in $V(\Phi_1,\Phi_2)$ allows mixing between the two
doublets, one diagonalises the charged and neutral scalar mass
matrices to obtain the physical states. There are altogether eight
mass eigenstates, three of which become the longitudinal components
of the $W^{\pm}$ and $Z$ gauge bosons.  Of the remaining five, there is a 
mutually conjugate pair of charged scalars ($H^{\pm}$), two neutral
scalars ($H, h$) and a neutral pseudoscalar ($A$), when there is no
CP-violation. Otherwise, a further mixing occurs between ($H, h$) and $A$. 
The compositions of the mass eigenstates $H$ and $h$ depend on the mixing angle $\alpha$.

In the absence of CP-violation, the squared masses of 
these physical scalars and the mixing angle $\alpha$ can be expressed as \cite{Gunion:2002zf},
\besub
\bea
\label{e:masq}
m_{A}^2 &=& \frac{m_{12}^2}{s_{\beta} c_{\beta}}-\frac{1}{2}v^2\left(2\lambda_5+\frac{\lambda_6}{t_{\beta}}+\lambda_7 t_{\beta}\right), \\
\label{e:mHpmsq}
m_{H^{\pm}}^2 &=& m_{A}^2+\frac{1}{2}v^2\left(\lambda_5-\lambda_4\right),\\
\label{e:mhsq}
m_{h}^2 &=& \frac{1}{2}\left[(A+B)-\sqrt{(A-B)^{2}+4 C^{2}}\right],\\
\label{e:mHsq}
m_{H}^2 &=& \frac{1}{2}\left[(A+B)+\sqrt{(A-B)^{2}+4 C^{2}}\right],\\
\label{e:tan2al}
\tan 2\a &=& \frac{2 C}{A-B},
\eea
\label{e:scalarmasses}
\eesub
where we have defined,
\besub
\bea
A &=& m_{A}^2 s_{\beta}^{2}+v^{2} (\lambda_1 c_{\beta}^{2}+\lambda_5 s_{\beta}^{2}+2 \lambda_6 s_{\beta} c_{\beta}),\\
B &=& m_{A}^2 c_{\beta}^{2}+v^{2} (\lambda_2 s_{\beta}^{2}+\lambda_5 c_{\beta}^{2}+2 \lambda_7 s_{\beta} c_{\beta}),\\
C &=& -m_{A}^2 s_{\beta} c_{\beta} +v^{2} \left[(\lambda_3+\lambda_4) s_{\beta} c_{\beta}+\lambda_6 c_{\beta}^{2}+\lambda_7 s_{\beta}^{2}\right].
\eea
\eesub
Furthermore, the interactions of the various charged and neutral scalars to the up- and down-type
fermions are functions of $\alpha$ and $\beta$. Their detailed forms in different 2HDM scenarios,
including the Type-II model adopted here for illustration, can be found in the literature \cite{Branco:2011iw}. 
 
\section{Theoretical and experimental constraints}
Next, we subject the Type-II 2HDM using various theoretical and
experimental constraints (though the most binding ones are often irrespective
of the specific type of 2HDM). It should be remembered at the outset that 
the most general $\mathbb{Z}_2$ violating 2HDM has seven quartic couplings, 
namely, $\lambda_{i}~(i=1,\ldots,7)$,  in addition to $\tan \beta$ and $m_{12}$,
totalling to nine free parameters. Though such a nine-dimensional parameter is
\emph{prima facie} large enough to accommodate any phenomenology, the set of
constraints under consideration below can ultimately become quite restrictive.

We discuss the theoretical constraints in subsection 3.1, and take up
the experimental/phenomenological ones in the subsequent subsections.
It should be noted that the parameter space is being constrained in
two distinct ways. Subsections 3.2 - 3.4 list essentially {\em
  low-energy} constraints which apply at the energy scale of the
subprocesses leading to Higgs production.  The various masses and
couplings get restricted by the requirement of satisfying them.
However, while such a strategy is valid for the discussion of
subsection 3.1 as well, we additionally require the conditions laid
down there to hold at various high scales, too. This not only
restricts the low-energy parameters more severely, but also answers
the main question asked in this paper, namely, to what extent the
2HDM can be deemed `ultraviolet complete'.

\subsection{Perturbativity, unitarity and vacuum stability}
For the 2HDM to behave as a perturbative quantum field theory at any
given scale, one must impose the conditions $\lvert \lambda_{i} \rvert
\leq 4\pi~(i=1,\ldots,7)$ and $ \lvert y_{i} \rvert \leq
\sqrt{4\pi}~(i=t,b,\tau)$ at that scale\footnote{The conditions are
  slightly different for the two types of couplings. The reason
  becomes clear if we note that the perturbative expansion
  parameter for $2 \rightarrow 2$ processes driven by the quartic
  couplings is $\lambda_i$. The corresponding parameter for
  Yukawa-driven scattering processes is $|y_i|^2$}.  On applying such
conditions, one implies upper bounds on the values of the couplings at
low as well as high scales.

  Next, we impose the more stringent condition of unitarity on the tree-level
  scattering amplitudes involving the scalar degrees of freedom. In a
  model with an extended scalar sector, the scattering amplitudes are
  taken between various two-particle states constituted out of 
  the fields $w_{i}^{\pm}$, $h_{i}$ and $z_{i}$ corresponding to the parametrisation of Eq.~\ref{e:doublet}. Maintaining this, there will be neutral two-particle states (e.g., $w_{i}^{+}w_{j}^{-},~ h_{i}h_{j},~ z_{i}z_{j},~ h_{i}z_{j}$) as well as singly charged two-particle states (e.g., $w_{i}^{\pm}h_{j},~ w_{i}^{\pm}z_{j}$).
The various two particle
  initial and final states give rise to a \emph{$2 \rightarrow 2$
    scattering matrix} whose elements are the lowest order partial
  wave expansion coefficients in the corresponding amplitudes.  The
  method used by Lee, Quigg and Thacker (LQT) \cite{Lee:1977eg} 
  prompts us to consider the eigenvalues of this 
  two-particle scattering matrix \cite{Ginzburg:2005dt,Gorczyca:2011he,Bhattacharyya:2013rya}. These eigenvalues, 
labelled as $a_i$, should satisfy the condition $|a_i| \leq 8 \pi$. Again,
these conditions apply to high scales as well, if we expect perturbativity
to hold.

When the quartic part of the scalar
potential preserves CP \cite{Grzadkowski:2013rza,Basso:2013wna} and $\mathbb{Z}_2$ symmetries, the LQT
eigenvalues are discussed in
\cite{Kanemura:1993hm,Akeroyd:2000wc,Horejsi:2005da}. For 
$\lambda_6, \lambda_7 = 0$, we follow the procedure 
and notation of \cite{Kanemura:1993hm} and \cite{Horejsi:2005da}. However, the matrices for coupled-channel analysis including $\l_6$ and $\l_7$ are derived by us (see $\mathcal{M}_{NC}$ and $\mathcal{M}_{CC}$ in Appendix~\ref{ss:LQT}). The general formulae including $\lambda_6, \lambda_7$, 
are given in Appendix \ref{ss:LQT}.

The condition to be taken up next is that of vacuum stability.
For the scalar potential of a theory to be stable, it must be bounded
from below in all directions. This condition is threatened if the quartic part
of the scalar potential, which is responsible for its behaviour at large
field values, turns negative. Avoiding such a possibility up to any given scale
ensures vacuum stability up to that scale. The issue of vacuum stability in context of
a 2HDM has been discussed in detail in \cite{Sher:1988mj,Nie:1998yn,Ferreira:2009jb,Kannike:2012pe,Barroso:2013awa}

The 2HDM potential has eight real scalar fields. By studying the
behaviour of the quartic part of its scalar potential along different
field directions, one arrives at the following conditions
\cite{Ferreira:2004yd,Branco:2011iw}, 
\besub 
\bea
\label{e:vsc1}
\rm{vsc1}&:&~~~\lambda_{1} > 0 \\
\label{e:vsc2}
\rm{vsc2}&:&~~~\lambda_{2} > 0 \\
\label{e:vsc3}
\rm{vsc3}&:&~~~\lambda_{3} + \sqrt{\lambda_{1} \lambda_{2}} > 0 \\
\label{e:vsc4}
\rm{vsc4}&:&~~~\lambda_{3} + \lambda_{4} - |\lambda_{5}| + \sqrt{\lambda_{1} \lambda_{2}} > 0 \\
\label{e:vsc5}
\rm{vsc5}&:&~~~\frac{1}{2}(\lambda_1+\lambda_2)+\lambda_{3} + \lambda_{4} +
\lambda_{5}-2|\lambda_6 + \lambda_7 | > 0 
\eea
\label{eq:vsc}
\eesub

The reader is reminded that the above conditions indicate a stable 
electroweak vacuum and not a metastable one. 
The couplings in the general $\mathbb{Z}_2$ violating Type-II 2HDM
evolve from a low scale to a high scale according to a set of
renormalisation group (RG) equations listed in the Appendix \ref{ss:RGE}.  If one
proposes the UV cut-off scale of the model to be some $\Lambda_{UV}$,
it might so happen that the couplings grow with the energy scale and
hit the Landau pole before $\Lambda_{UV}$. A \emph{second}, still
unacceptable, possibility is that of the LQT eigenvalues crossing their
unitarity limits. The RG evolution of the
2HDM couplings has been recently studied in \cite{Bijnens:2013rd,Chakraborty:2014oma}. Finally, the stability conditions can get violated
below $\Lambda_{UV}$, making the scalar potential unbounded from
below.  All these problems are avoided if one postulates that all of
the conditions laid down above are valid up to $\Lambda_{UV}$, which
marks the maximum energy up to which the 2HDM can be valid without the
intervention of any additional physics.

\subsection{Higgs mass constraints}
The spectrum of a generic 2HDM consists of a charged scalar, a CP-odd
neutral scalar and two CP-even neutral scalars. Since the LHC has
observed a CP-even neutral boson around 125 GeV, we allow only those
regions in the parameter space for which $h$, the lighter neutral
scalar, lies in the mass range 124.53-126.18 GeV which is within 2$\sigma$ error limits following \cite{Aad:2014aba}. In
addition, the charged scalar is required to have a mass greater than
315 GeV due to low energy constraints, coming mainly 
from $b \rightarrow s \gamma$\cite{Mahmoudi:2009zx,Deschamps:2009rh}. 
The benchmark points used by us are also consistent with $B \rightarrow \tau \nu_{\tau}$, $B_{s}\to \mu^{+}\mu^{-}$ and $B^0$-$\overline{B^0}$ mixing~\cite{Beringer:1900zz,Deschamps:2009rh}.

\subsection{Oblique parameter constraints} 
The presence of an additional $SU(2)$ doublet having a hypercharge $Y = 1$ modifies the electroweak oblique parameters \cite{Peskin:1991sw}. It is to be noted that since the couplings of the fermions to gauge bosons remain unaltered even after the introduction of the second doublet, all the additional contributions come from the scalar sector of the 2HDM. The oblique parameters can be decomposed as,
\besub
\bea
S = S_{SM} + \Delta S \\
T = T_{SM} + \Delta T,
\eea
\eesub
where $S_{SM}$ and $T_{SM}$ denote the Standard Model (SM) contributions and $\Delta S$ and $\Delta T$ denote any new physics effect.
The central value is the contribution coming from the standard model with the reference values $m_{h,\rm{ref}} = 125.0$ GeV and $M_{t,\rm{ref}} = 173.1$ GeV where $M_t$ denotes the pole mass of the top quark. 
The expressions for $\Delta S$ and $\Delta T$ for a general 2HDM can be found in \cite{He:2001tp, Grimus:2008nb, Haber:2010bw, Funk:2011ad}. The corresponding bounds we have used are $|\Delta S| < 0.11 $ and $|\Delta T| < 0.13$ \cite{Baak:2014ora}.
The splitting amongst the scalar masses affects the $T$ parameter, which is linked to the custodial $SU(2)$ symmetry. Typically for $m_{12} = 0$, $T$ prevents large mutual splitting among states other than the lightest neutral scalar. For $m_{12} \neq 0$, the scalars other than the light neutral one have masses $ \sim m_{12}$. As $m_{12}$ is increased, the masses approach the decoupling limit, and in that case, the oblique electroweak constraints are naturally satisfied, as the 2HDM approaches the SM in that case. The consistency with these parameters has nevertheless been explicitly ensured at each allowed point of the parameter space. 

\subsection{Collider constraints}
Apart from the theoretical constraints discussed above, we also strive to find the region of parameter space of a 2HDM allowed by the recent Higgs data. The  ATLAS~\cite{ATLAS:2012dsy, TheATLAScollaboration:2013lia, Aad:2013wqa} and CMS~\cite{CMS:ril} collaborations have measured the production cross section
 for a $\sim$125 GeV Higgs multiplied by its branching ratios
 to various possible channels. In our case, since the underlying theory is a 2HDM, all the cross sections and decay widths get modified compared to the corresponding SM values. For example, the production cross section of the light neutral Higgs through gluon fusion will get rescaled in the case of a 2HDM due to the fact that the fermionic couplings of  the 125 GeV Higgs are now changed with respect to the SM values by appropriate multiplicative factors. Similarly, the loop induced decay $h \to \gamma \gamma$ will now draw an additional contribution from the charged scalars. Some recent investigations in this area can be found in \cite{Coleppa:2013dya, Ferreira:2011aa, Ferreira:2013qua, Drozd:2012vf,Cheon:2012rh, Drozd:2013nta,Krawczyk:2013gia,Wang:2013sha, Dumont:2014wha,  Celis:2013rcs, Celis:2013ixa, Eberhardt:2013uba, Kanemura:2014bqa}. Also, model-independent analysis of the data, which impose constraints on non-SM couplings of the scalar discovered, have to allow such contributions~\cite{Banerjee:2012xc,Corbett:2012dm,Espinosa:2012ir,
 Giardino:2012ww,Klute:2012pu,Azatov:2012rd}.
In order to check the consistency of a 2HDM with the measured rates in various channels,
we theoretically compute the signal strength $\mu^{i}$ for the $i$-th channel using the relation:
\be
\mu^{i} = \frac{R_{\rm{prod}} \times R_{\rm{decay}}^i}{R_{\rm{width}}}~.
\ee    
Here $R_{\rm{prod}}$, $R_{\rm{decay}}^i$ and $R_{\rm{width}}$ denote respectively the ratios of the theoretically calculated production cross section, the decay rate to the $i$-th channel and the total decay width for a $\sim$125 GeV Higgs to their corresponding SM counterparts. Thus, our analysis strategy is to generate a region in parameter space allowed by the constraints coming from vacuum stability, perturbative unitarity and electroweak precision data. We subsequently compute $\mu^{i}$ for each point in that allowed region and compare them to the experimentally measured signal strengths, $\hat{\mu}^i$, supplied by the LHC. This exercise carves out a sub-region, which is allowed by the recent Higgs data, from the previously obtained parameter space. We have implemented the Runge-Kutta algorithm to solve the RG equations through our own code. The oblique parameters and the signal strengths to various channels have been computed using standard formulae available in the literature. Moreover, the consistency of the obtained results have been checked using the public code 2HDMC~\cite{Eriksson:2009ws} at various parameter points.

For our numerical analysis, we have taken gluon fusion to be the dominant production mode for the SM-like Higgs.\footnote{While other channels such as vector boson fusion (VBF) and associated Higgs production with W/Z (VH) have yielded data in the 8 TeV run, the best fit signal strengths are still dominated by the gluon fusion channel. Here our primary task is to check the high scale validity of the 2HDM. In that approximation, the K-factors in $\sigma$ and $\sigma_{SM}$ are taken to be the same.} As for the subsequent decays of $h$, we have considered all the decay channels mentioned in Table \ref{tab:tab1}. We use $1\s$ allowed ranges of $\hat{\mu}^{i}$. 
 
\begin{table}[h]
\centering
\begin{tabular}{|c|c|c|c|}
\hline
Channel                                            & Experiment & $\hat{\mu}$            & Energy in TeV  (Luminosity in fb$^{-1}$) \\ \hline \hline
\multirow{2}{*}{$h \rightarrow \gamma \gamma$}     & ATLAS      & $1.55^{+0.33}_{-0.28}$ & $7~(4.8)$ + $8~(20.7)$                   \\ \cline{2-4} 
                                                   & CMS        & $1.13^{+0.24}_{-0.24}$ & $7~(5.1)$ + $8~(19.6)$                   \\ \hline
\multirow{2}{*}{$h \xrightarrow{Z Z^{*}} 4l$}      & ATLAS      & $1.43_{-0.35}^{+0.40}$ & $7~(4.6)$ + $8~(20.7)$                   \\ \cline{2-4} 
                                                   & CMS        & $1.00_{-0.29}^{+0.29}$ & $7~(5.1)$ + $8~(19.7)$                   \\ \hline
\multirow{2}{*}{$h \xrightarrow{W W^{*}} 2l 2\nu$} & ATLAS      & $0.99_{-0.28}^{+0.31}$ & $7~(4.6)$ + $8~(20.7)$                   \\ \cline{2-4} 
                                                   & CMS        & $0.83_{-0.21}^{+0.21}$ & $7~(4.9)$ + $8~(19.4)$                   \\ \hline
\multirow{2}{*}{$h \rightarrow b \bar{b}$}         & ATLAS      & $0.20_{-0.60}^{+0.70}$ & $7~(4.7)$ + $8~(20.3)$                   \\ \cline{2-4} 
                                                   & CMS        & $0.91_{-0.49}^{+0.49}$ & $7~(5.1)$ + $8~(18.9)$                   \\ \hline
\multirow{2}{*}{$h \rightarrow \tau \bar{\tau}$}   & ATLAS      & $1.4_{-0.40}^{+0.50}$  & $8~(20.3)$                               \\ \cline{2-4} 
                                                   & CMS        & $0.91_{-0.27}^{+0.27}$ & $7~(4.9)$ + $8~(19.7)$                   \\ \hline \hline
\end{tabular}
\caption{The signal strengths in various channels with their 1$\sigma$ uncertainties.}
\label{tab:tab1}
\end{table}

\section{Results with exact discrete symmetry}
In this section, we set out to obtain the allowed parameter space of a  Type-II 2HDM having an \emph{exact} $\mathbb{Z}_2$ symmetry consistent with the various theoretical and collider constraints described above. In this particular case, one naturally has $m_{12}=0$, $\lambda_6,\lambda_7 = 0$. Thus, we scan over the quartic couplings $ \lambda_{i} ~(i=1,\ldots,5)$ within their perturbative limits ($\l_{1,2}\in [0,4\pi ]~\rm{and}~\l_{3,4,5}\in [-4\pi, 4\pi]$) and allow them to evolve from a low scale to a higher scale, designated by $\L_{UV}$. 
The RG equations for the evolution of all the 2HDM couplings are listed in Appendix \ref{ss:RGE}. In our analysis, the scale from which the evolution starts, has been chosen to be the top quark pole mass $M_t = 173.1$ GeV. This pins down the values of the Yukawa couplings at that scale through the relations $y_t(M_t)=\sqrt{2} m_t(M_t)/v_2$ and $y_i(M_t)=\sqrt{2} m_i(M_t)/v_1$ for $i = b~{\rm and}~\tau$. Here $m_j(M_t)$ refers to the \emph{running} mass of the $j$-th fermion at the scale $M_t$ 
in $\overline{MS}$ scheme. We choose $m_t(M_t),~ m_b(M_t)$ and $m_{\tau}(M_t)$ to be 163.30, 4.20 and 1.77 GeV respectively~\cite{Moch:2010rh, Langenfeld:2010zz}.  

We obtain the allowed values of $ \lambda_{i}(M_t) ~(i=1,\ldots,5)$ which, in course of evolution towards $\L_{UV}$, satisfy all the constraints of perturbativity, unitarity and vacuum stability at all intermediate scales. We choose $\L_{UV} = 1$ TeV and $\tan\b = 2$ as an appropriate benchmark. 

\begin{figure}[!htbp]
\begin{center}
\includegraphics[scale=0.45]{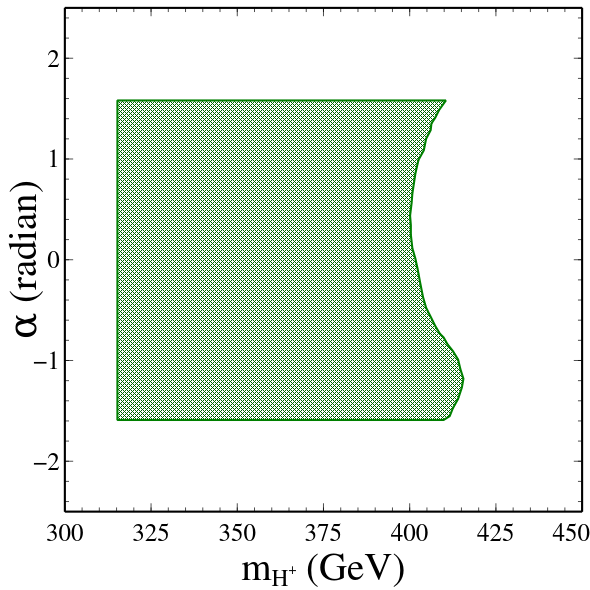}~~~~~~~~
\includegraphics[scale=0.45]{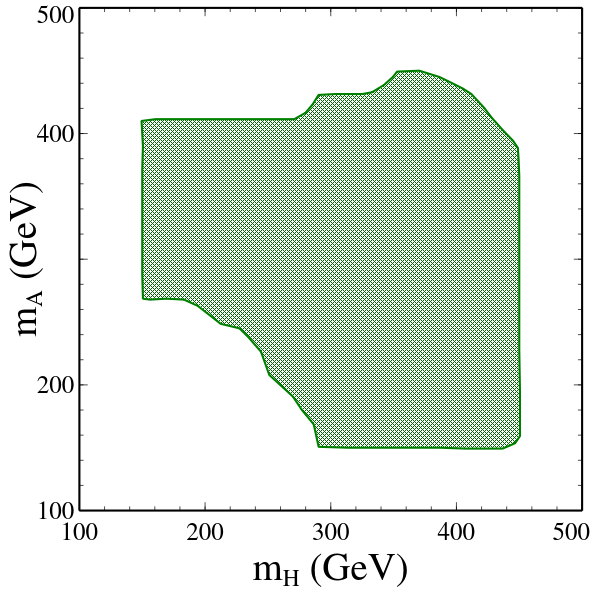}
\caption{Theoretically allowed parameter spaces at $\L_{UV} = 1$ TeV, $\tan\beta = 2$ and $m_{12} = 0$  GeV for $M_{t}$ = 173.1 GeV. The region in the figure on the left is allowed concomitantly with that in the figure on the right.}
\label{f:diffmt_E3}
\end{center}
\end{figure}

\begin{figure}[!htbp]
\begin{center}
\includegraphics[scale=0.5]{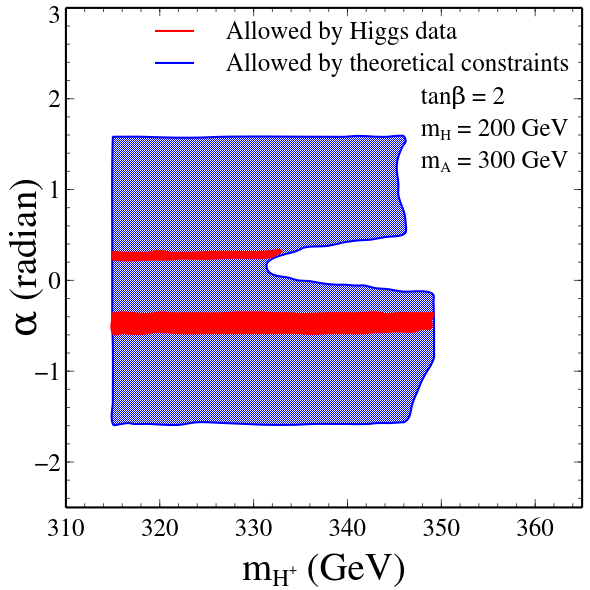}~~~~~~~~
\includegraphics[scale=0.5]{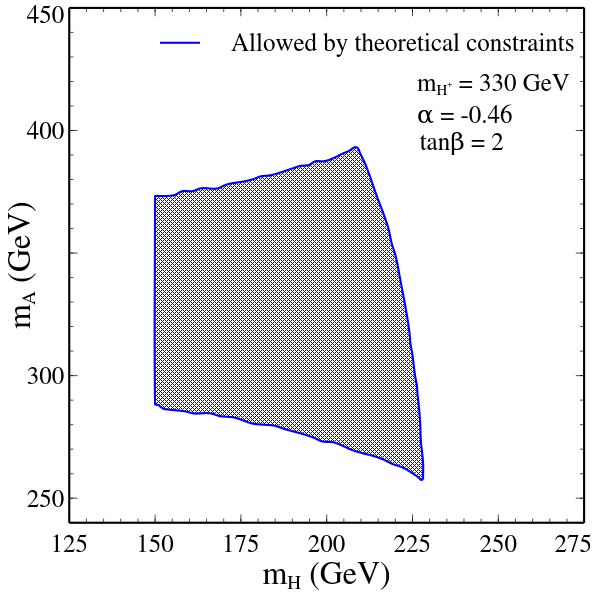}
\caption{Region allowed in the $m_{H^\pm}$-$\alpha$ and $m_{H}$-$m_{A}$ planes, by the theoretical constraints and the recent Higgs data. In each case, the chosen benchmark values of the two other parameters are given in the legend.}
\label{f:mhplal_BP}
\end{center}
\end{figure}

We display our scan results as allowed regions of parameter space in $m_H-m_A$ plane as well in the $m_{H^\pm}-\alpha$ plane as shown in Fig.~\ref{f:diffmt_E3}. The oblique parameters play a role in restricting the splitting between the masses. Moreover, demanding perturbative unitarity and vacuum stability up to the TeV scale causes the allowed region to shrink further. In other words, for any value of the masses not within the allowed region, the quartic couplings are such that, if they are used as initial conditions in the RG equations, they would violate perturbative unitarity or vacuum stability below the TeV scale. For example, vacuum stability up to the TeV scale puts an upper bound on $|\l_5|$ which in turn translates into an upper bound on $m_A$ (see Eqs.~\ref{e:masq} and~\ref{e:vsc4}). 
The mixing angle $\alpha$ gets further constrained by the recent Higgs data. Since, values of $m_H$ and $m_A$ chosen do not play a role in modifying the Higgs signal strengths, we choose a benchmark $m_H$ = 200 GeV and $m_A$ = 300 GeV and project the allowed region in the $m_{H^\pm}-\alpha$ plane, shown in Fig.~\ref{f:mhplal_BP}.   

Recent data indicate that $M_{t}$ the top quark pole mass is $[173.07\pm 0.52\pm 0.72]$ GeV~\cite{Beringer:1900zz}\footnote{We have used the allowed range
of the top quark pole mass as given in the above reference. The allowed
range changes slightly, according to the most recent result~\cite{Abazov:2014dpa}}. Different values for $M_t$ (within the allowed band) necessarily alter the running masses as well. However, choosing different values of the top quark mass does not cause any noticeable change to
the allowed region in the parameter space of scalar masses and mixing angle. Since in this case, the RG running of the quartic couplings takes place over a relatively shorter length of energy scale, i.e., from the electroweak scale to 10 TeV, the evolution trajectories corresponding to different values of the top mass do not diverge apart from each other. For example, it has been checked that the allowed space in terms of masses, where we have used $M_t = 173.1$ GeV, remains almost identical if $M_t$ takes any value between 171.0 and 175.2 GeV.

We thus can say that, in case of exact $\mathbb{Z}_2$ symmetry, the uncertainty in the top quark mass measurement has almost no bearing on the allowed region of the parameter space. This result alerts us to a more important one that we obtain in the next sections, namely, the high scale validity of the 2HDM irrespective of the measured value of the top quark mass. 

For $\beta-\alpha=\pi/2$, the 2HDM couplings of the 125 GeV Higgs to fermions and gauge bosons are the same as the SM ones. In that case, the Higgs signal strengths to various channels should match with the corresponding SM ones. Fig.~\ref{f:mhplal_BP} shows an allowed band around $\alpha = \beta - \pi/2 = -0.46$ thus validating this observation.  
Over the entire region marked with red in Fig.~\ref{f:mhplal_BP}, $\cos(\beta-\alpha)$ is very small. As a result, $m_H$ = 200 GeV is not ruled out by the LHC data, since the $ZZ$ and $WW$ decay modes of $H$ are suppressed.
To illustrate the RG running of the various couplings, the vacuum stability conditions and the LQT eigenvalues, we choose the following initial conditions,
\be
\l_1(M_t) = 1.33,~\l_2(M_t) = 0.90,~\l_3(M_t) = 4.08,~\l_4(M_t) = -2.13,\mbox{and}~\l_5(M_t) = -1.79~.
\label{e:bc1}
\ee

This choice of boundary conditions for our illustration is aimed at
keeping $\lambda_1$ as low as possible, with $m_h$ in the right range. We want to show that even with such a choice, the theory violates perturbativity and unitarity below 10 TeV. Thus the impossibility of this 2HDM with $m_{12} = 0$ at high scale gets established.
\begin{figure}[!ht]
\begin{center}
    \subfloat[ \label{sf:t2m0l}]{%
      \includegraphics[scale=0.3]{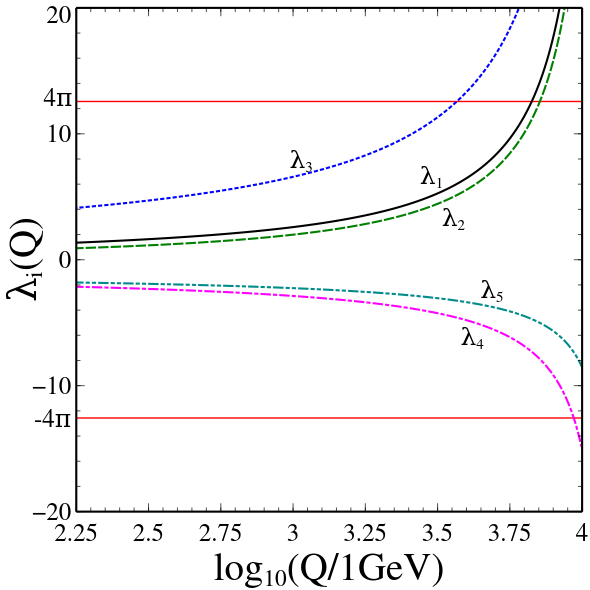}
    }~~~~
    \subfloat[\label{sf:t2m0LQT}]{%
      \includegraphics[scale=0.3]{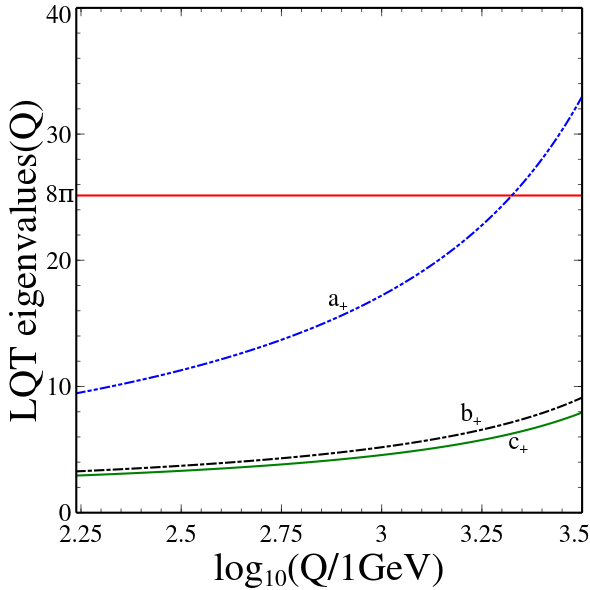}~~~~
    }
    \subfloat[\label{sf:t2m0vsc}]{%
      \includegraphics[scale=0.3]{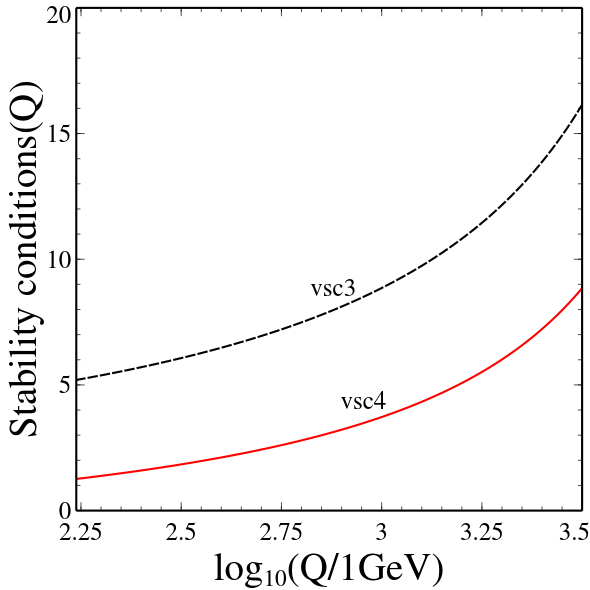}
    }    
    \caption{RG running of $\lambda_i$, the LQT eigenvalues and the stability conditions with the energy scale for $\tan\beta = 2$ and $m_{12} = 0$. The horizontal
    lines in the leftmost figure (\ref{sf:t2m0l}) denote the perturbative limit and unitarity limit in the second figure (\ref{sf:t2m0LQT}). Also $a_+$, $b_+$ and $c_+$ in the second figure (\ref{sf:t2m0LQT}) are the LQT eigenvalues explained in the Appendix~\ref{ss:LQT}. In the rightmost figure (\ref{sf:t2m0vsc}), vsc3 and vsc4 represent the two stability conditions that are defined in Eq.~\ref{eq:vsc}.}
    \label{f:t2m0}
\end{center}
\end{figure}
Fig.~\ref{f:t2m0} describes the RG running of $\l_i$ with the aforementioned low values as boundary conditions. These values correspond to, 
\begin{gather}
m_h = 125.44 {~\rm GeV}, ~m_H = 210.00 {~\rm GeV}, ~m_{H^{\pm}} = 345.00 {~\rm GeV},\nonumber \\
m_A = 330.00 {~\rm GeV},~\alpha = 0.95{~\rm radian} ~.
\end{gather}

 which is an allowed point in the parameter space, as shown in Fig.~\ref{f:mhplal_BP}. Since $\l_1$ starts evolving from rather large a value, it rises steeply with the energy scale. For the value of $\tan\beta$ taken here, $y_{b}$ and $y_{\tau}$ have small initial values, and hence, they do not slow the evolution curve down to any appreciable extent (see Eq.~\ref{e:lam1RG}). On the other hand, $y_t$, being the dominant Yukawa coupling in this case, prevents $\l_2$ to rise as sharply as $\l_1$. The LQT eigenvalues (see Eq.~\ref{e:LQTeval}) $a_{+}$, $b_{+}$ and $c_{+}$ evolve in a manner as shown in Fig~\ref{sf:t2m0LQT}. Also, the stability conditions remain positive during the course of evolution, as shown in Fig.~\ref{sf:t2m0vsc}. A different initial condition which has a higher value of $\l_1(M_t)$ for instance, would 
lead to steeper evolution trajectories for the couplings. Hence, the overall conclusions regarding high-scale validity of this scenario would not alter.   
  
This leads to the observation that the various $\l_i$ become non-perturbative below a scale of 10 TeV. Also, it is seen that the LQT eigenvalue $a_+$ hits the unitarity limit faster than the quartic couplings hit their perturbative limits. Thus, this example illustrates the interplay among \emph{perturbativity} and \emph{unitarity} in determining the UV fate of this scenario and it appears that unitarity often proves stronger as a constraint than perturbativity. It should also be 
noted that all plots in Figs.~\ref{f:diffmt_E3} and \ref{f:mhplal_BP} use $\tan\beta = 2$. This is because the quartic
couplings cannot be kept in their perturbative limits for $\tan\beta\ge3$. A wider scan over the parameter space corresponds to this observation. 
Also, one can generally conclude that in order to push the UV limit of 2HDM to higher scales, one must look beyond the \emph{exact} $\mathbb{Z}_2$ symmetric case.

\section{Results with softly broken discrete symmetry}

\begin{figure}[!htbp]
\begin{center}
\includegraphics[scale=0.3]{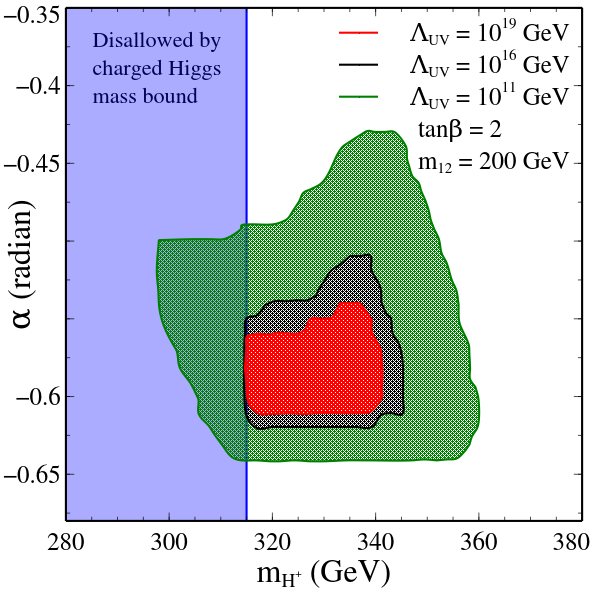}~~~~~~
\includegraphics[scale=0.3]{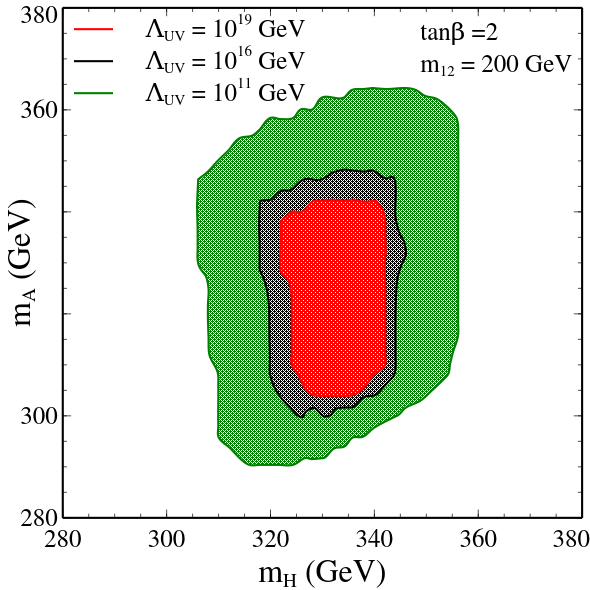}\\
\vspace*{0.3cm}
\includegraphics[scale=0.3]{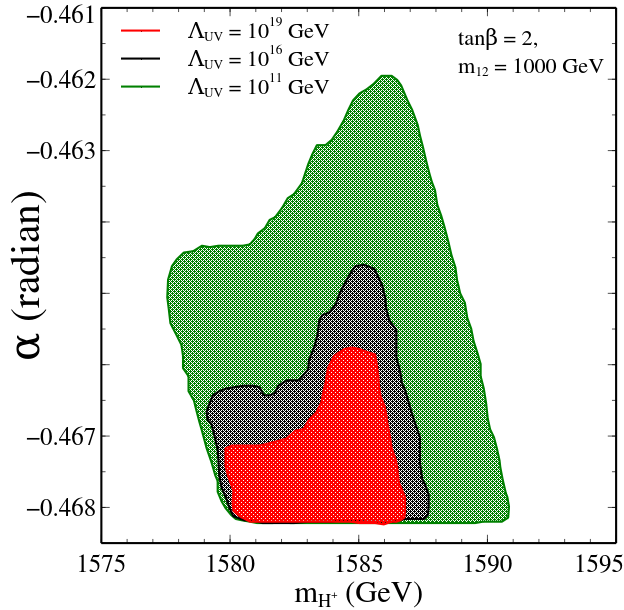}~~~~~~
\includegraphics[scale=0.3]{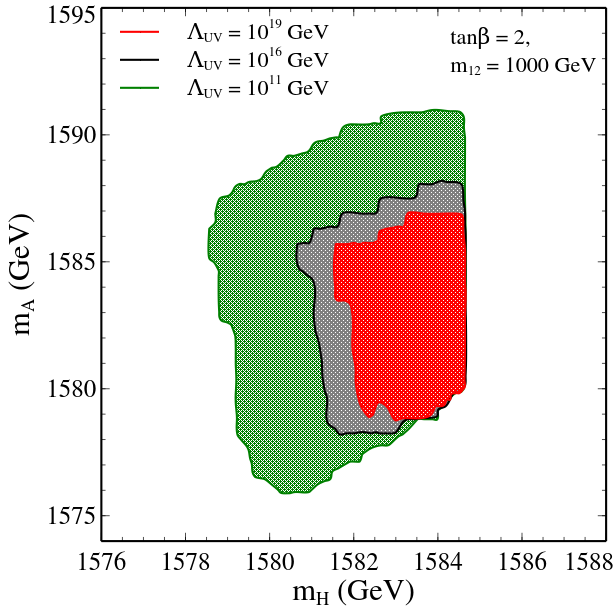}\\
\vspace*{0.3cm}
\includegraphics[scale=0.3]{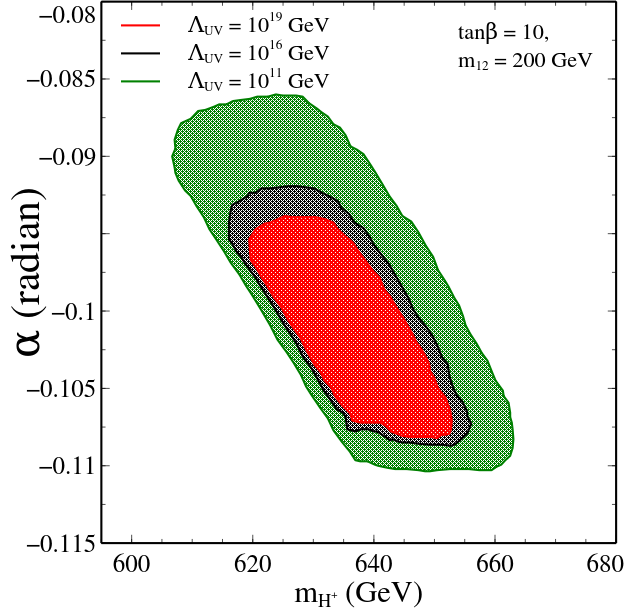}~~~~~~
\includegraphics[scale=0.3]{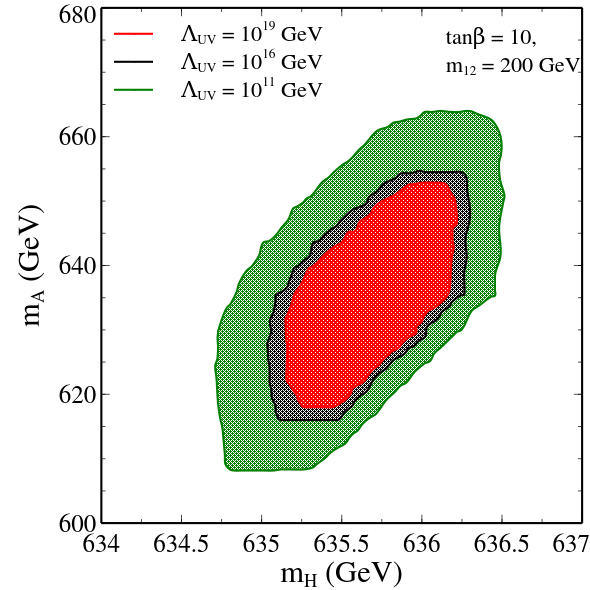}\\
\vspace*{0.3cm}
\includegraphics[scale=0.3]{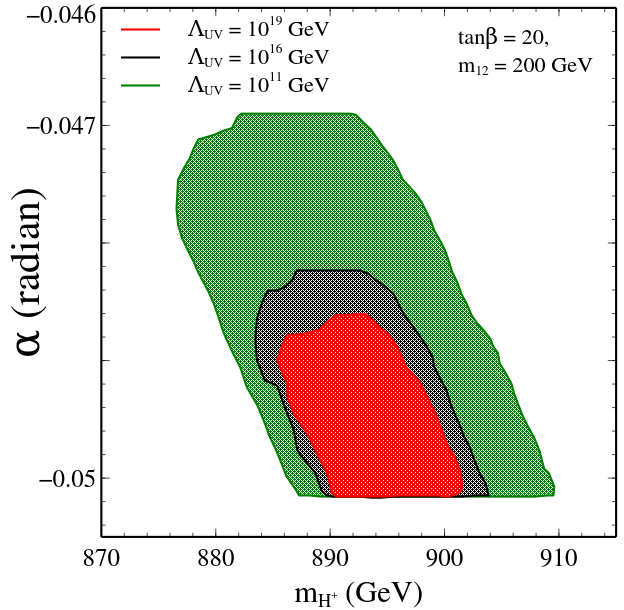}~~~~~~
\includegraphics[scale=0.3]{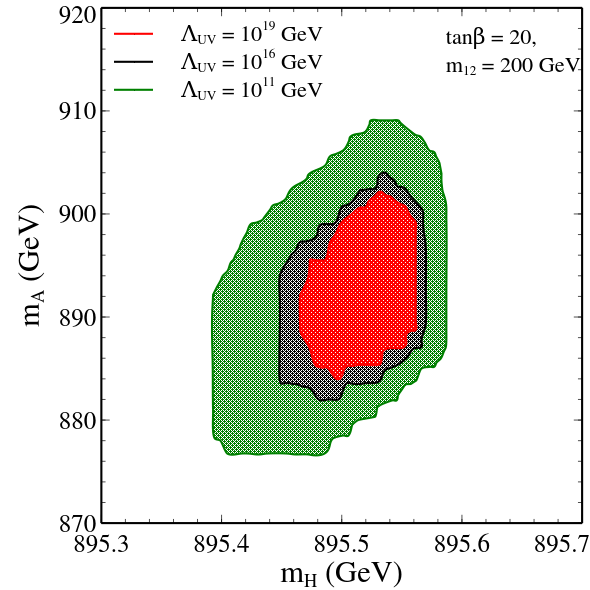}
\end{center}
\caption{The allowed parameter spaces in the soft $\mathbb{Z}_2$ breaking case for $\Lambda_{UV} = 10^{11}$ (green), $10^{16}$ (grey) and $10^{19}$ GeV (red). The $\tan\beta$ and $m_{12}$ values are shown in the plots. The shaded region (blue) in the top left figure denotes the exclusion coming from flavour constraints.}
\label{f:wo67E111619}
\end{figure}

\begin{figure}[!htbp]
\begin{center}
\includegraphics[scale=0.3]{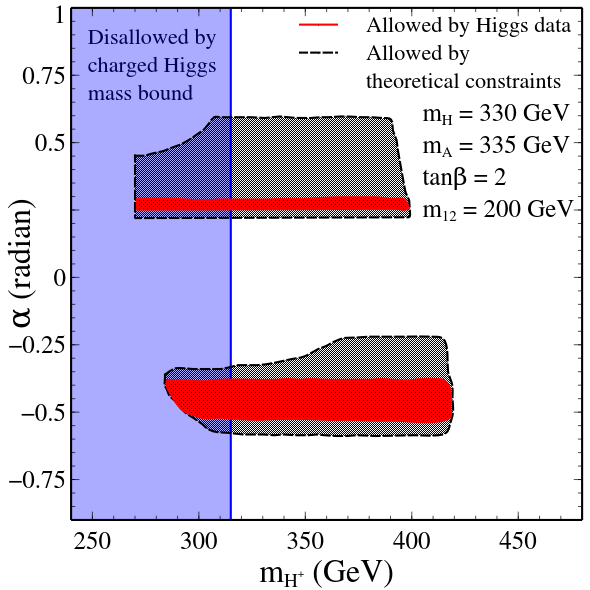}~~~~~~
\includegraphics[scale=0.3]{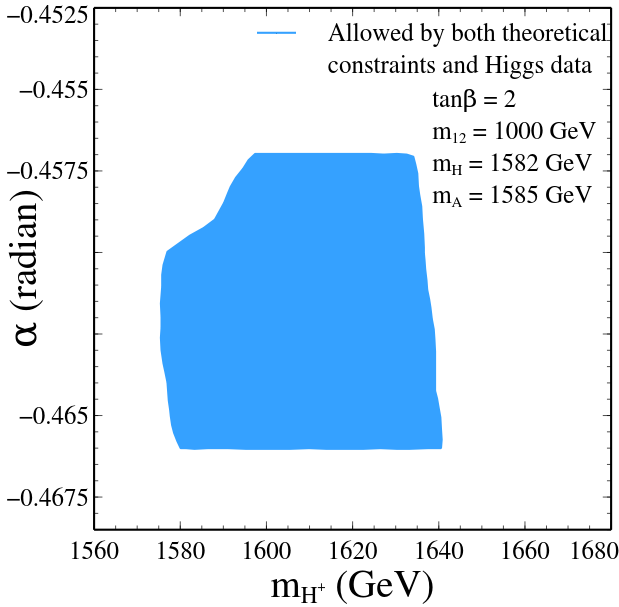}\\
\vspace*{0.3cm}
\includegraphics[scale=0.32]{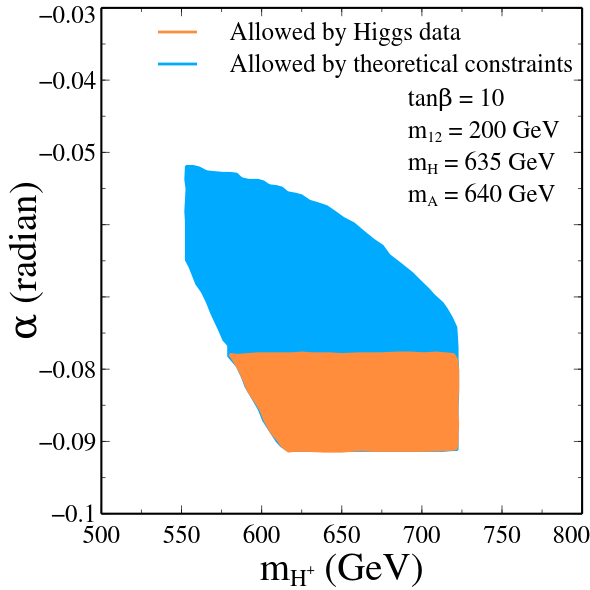}~~~~~~
\includegraphics[scale=0.32]{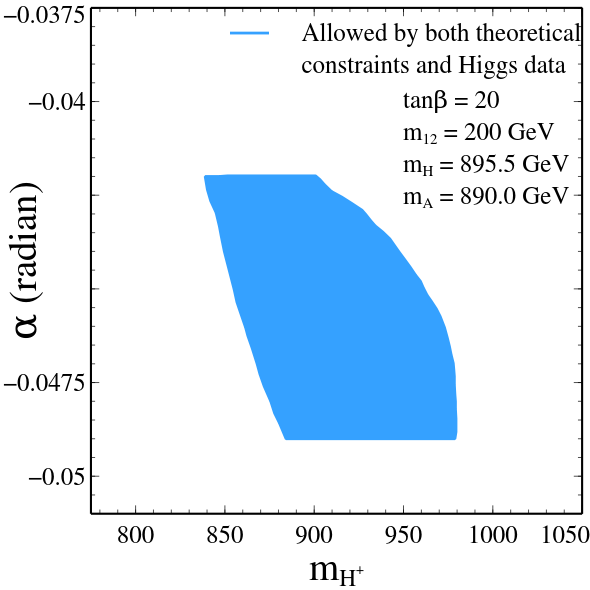}
\caption{Regions in the $m_{H^{\pm}}$-$\alpha$ plane allowed by the Higgs data in the soft $\mathbb{Z}_2$ breaking case. }
\label{f:wo67E3coll}
\end{center}
\end{figure}

This section illustrates the effects of the various constraints imposed on the model with non-zero $m_{12}$, i.e., in presence of a \emph{soft} $\mathbb{Z}_2$ symmetry violating term. The RG runnings of the various couplings in the model are just like the ones in \emph{exact} $\mathbb{Z}_2$ symmetric case, the only differences being in the expressions for the scalar masses as evident from Eq.~\ref{e:scalarmasses}. We scan the model parameter space and look for points which satisfy all the constraints listed in Sec. 3 up to $\Lambda_{UV} = 10^3, 10^{11}, 10^{16}, 10^{19}$ GeV in Figs.~\ref{f:wo67E111619}, and~\ref{f:wo67E3coll}. Validity of the model up to the TeV scale, Grand Unification scale, Planck scale as well as $\sqrt{M_{\rm Pl} M_{\rm TeV}}$ is addressed in this manner\footnote{Here $M_{\rm Pl}$ and $M_{\rm TeV}$ represents the Planck scale and TeV scale respectively. The benchmarks chosen are $\tan\beta = 2,10,20$ and $m_{12} = 200,1000$ GeV, which represent the electroweak and TeV scales. This also keeps the 2HDM spectrum within the ultimate reach of the LHC. Having $\tan\beta$ higher
than in the previous section is possible in this case, so long as $m_{12}$ is correspondingly large, thus generating an acceptable $m_h$. For $\L_{UV} = 10^{11}, 10^{16}, 10^{19}$ GeV, we project our allowed results as two dimensional contour plots in the $m_H-m_A$ and $m_{H^\pm}-\alpha$ planes. In each row, the plots in the left- and right-hand sides represent concomitantly allowed regions. This choice pins down the 2HDM parameter space in terms of the physically measurable observables.}  \\

The results in the figure show that as we go higher in $\L_{UV}$, the allowed parameter space shrinks. 
The splitting amongst the scalar masses becomes the narrowest at the Planck scale, albeit being dependent on the values of $\tan\beta$ and $m_{12}$. 

An inspection of the results so obtained shows that as $\Lambda_{UV}$ is pushed towards higher scales, the allowed parameter space shrinks, and finally at the Planck scale, it is most constrained. For example, for $\tan\beta = 10$, $m_{12} = 200$ GeV and $\L_{UV} = 10^{19}$ GeV, the masses (in GeV) are observed to lie in the following range,
\be
m_H \in [635,636],~m_{H^{\pm}} \in [619,652],~m_A \in [618,653]~.
\ee
We note here that though $m_{12}$ does not appear in the RG equations themselves, it indirectly puts constraints on $\l_i$ through the mass constraints.

\begin{figure}[!htbp]
\begin{center}
\includegraphics[scale=0.4]{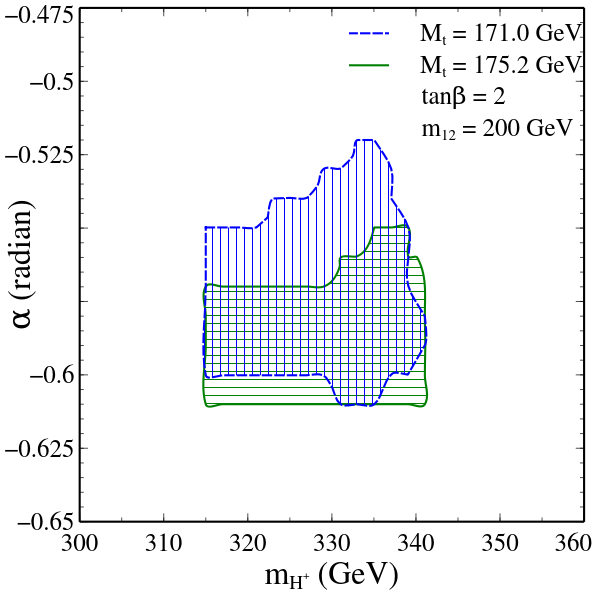}~~~~~~~~
\includegraphics[scale=0.4]{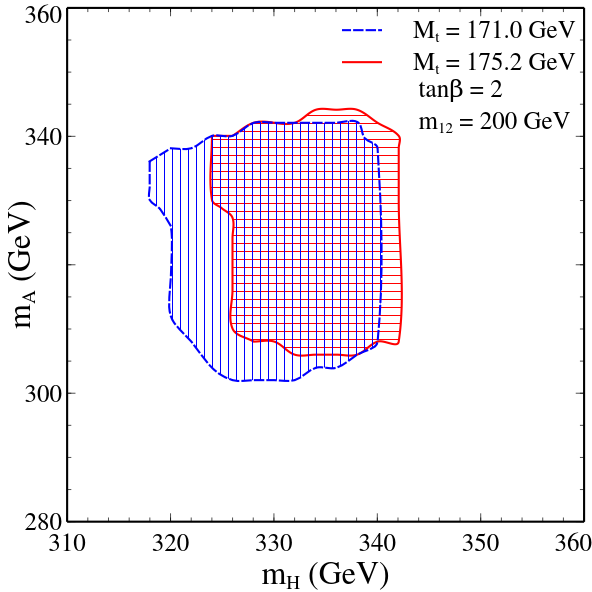}
\end{center}
\caption{A comparison of the allowed parameter spaces at $\L_{UV} = 10^{19}$ GeV, $\tan\beta = 2$ and $m_{12} = 1000$ GeV for two values of $M_{t}$, in the soft $\mathbb{Z}_2$ breaking case.}
\label{f:diffmt_E19}
\end{figure}

 Note that since $\tan\beta$ determines the initial conditions for the Yukawa couplings, it does affect the RG running of $\lambda_i$. Although $m_b(M_t)$ and $m_\tau(M_t)$ are small compared to $m_t(M_t)$, for a high $\tan\beta$, $y_b(M_t)$ could be comparable to $y_t(M_t)$ . This is the main motivation behind our choosing $\tan\beta = 20$. A change in the top quark mass is expected to modify the obtained parameter space to a considerable extent. This fact is illustrated in Fig.~\ref{f:diffmt_E19} where we choose $M_t = 171.0, 175.2$ GeV and highlight the difference in the parameter spaces so obtained.  In the subsequent sections, we keep $M_t = 173.1$ GeV.It may be argued that in determining the high scale validity of the model, the relatively less crucial role played here by the top quark mass is just due to the larger number of free parameters in the 2HDM scenario. While this is true in a sense, the analysis reported in Fig.~\ref{f:diffmt_E19} was still required for the following reason. To counter the downward evolution of $\l_2$ due to the top quark Yukawa coupling (see Eq.~\ref{e:lam2RG}), the participation of the other $\l_i$ plays a role. However, large values of these parameters may again violate perturbative unitarity, and in turn prevent one from extending the theory to high energy scales. The lesson to learn from  Fig.~\ref{f:diffmt_E19} is that valid regions in the parameter space can be found, which survive the above tug-of-war. Consequently, a Type-II 2HDM may hold true till the Planck scale without any additional new physics, even for high-end values of the top quark mass.

The impact of the recent LHC data on the parameter space already allowed by the theoretical constraints is shown in Fig.~\ref{f:wo67E3coll}. In this case, we pick up benchmark values of $m_H$ and $m_A$ suitably to avoid the direct search constraints. In addition, these benchmarks are chosen from a region satisfying the theoretical constraints up to the Planck scale. The 2HDM decay widths are sensitive to the mixing angles and the charged scalar mass and the collider constraints carve out a subregion in the $m_{H^{\pm}}-\alpha$ plane.

\begin{figure}[!htbp]
\begin{center}
\includegraphics[scale=0.3]{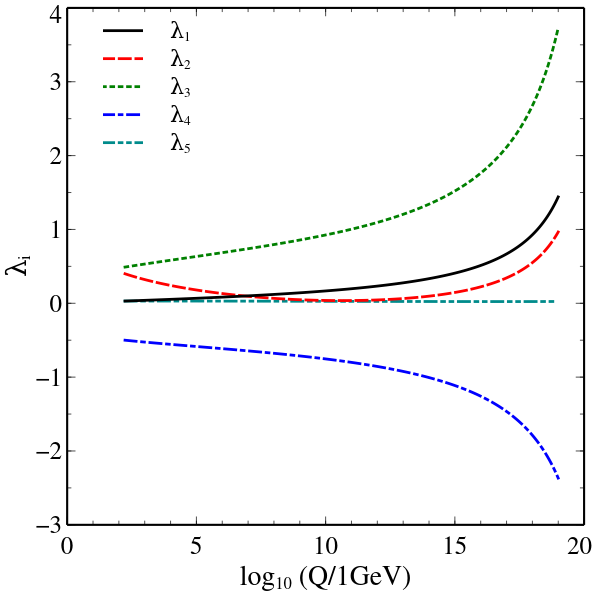}~~~~
\includegraphics[scale=0.3]{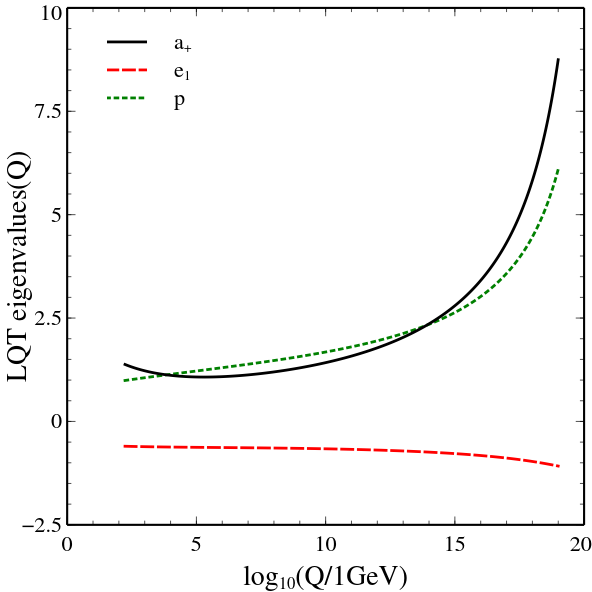}~~~~
\includegraphics[scale=0.3]{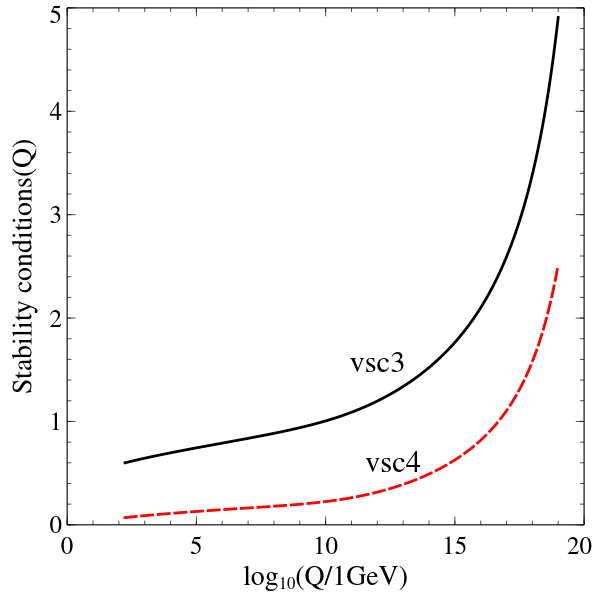}
\end{center}
\caption{RG running of $\lambda_i$, the LQT eigenvalues and the stability conditions with the energy scale for $\tan\beta = 2$  and $m_{12}=1000$ GeV in the soft $\mathbb{Z}_2$ breaking case.}
\label{f:E19running}
\end{figure}

The figure shows allowed bands around $\alpha = \beta - \pi/2$ in each case. Note that for scalar masses $\sim 1$ TeV or more, the entire region allowed by the theoretical constraints is favoured by the Higgs data. This is precisely due to the decoupling of the heavier degrees of freedom from the theory.
A small enough initial value of $\lambda_2$ causes $\lambda_2 (Q)$ to turn negative at some scale affecting the vacuum stability of the theory thereby. To illustrate the point better, we choose an initial condition,
\be
\l_1(M_t) = 0.03,~\l_2(M_t) = 0.39,~\l_3(M_t) = 0.49,~\l_4(M_t) = -0.50~\mbox{and}~\l_5(M_t) = 0.03,
\label{e:bc2}
\ee
 for the quartic couplings at $\tan\beta = 2$  and $m_{12}=1000$ GeV, out of the allowed set of couplings which obey all the imposed constraints up to the $\L_{UV} = 10^{19}$ GeV. These quartic couplings expressed in terms of the masses and the mixing angle become, 
\begin{gather}
m_h = 124.78{~\rm GeV}, ~m_H = 1582.31{~\rm GeV}, ~m_{H^{\pm}} = 1585.64{~\rm GeV},\nonumber \\
m_A = 1580.56{~\rm GeV},~\alpha = -0.466{~\rm radian}~.
\end{gather}
 We display the RG running of the $\lambda_i$, the stability conditions and the LQT eigenvalues in Fig.~\ref{f:E19running}. This choice of sample boundary conditions here is guided by a consideration complimentary to that
 of Fig.~\ref{f:t2m0}. Here we show that it is possible to identify points in the parameter space, which correspond to the quartic couplings avoiding any perturbativity, unitarity or vacuum stability constraints all the way up to the Planck scale.
As indicated in Fig.~\ref{f:E19running}, $a_{+}(Q)$ grows most sharply amongst the other LQT eigenvalues and hence violates unitarity just after crossing the Planck scale in this case. Thus it turns out that $|a_{+}(Q)| \leq 8\pi$ proves to be the strongest constraint in determining an upper bound on $|\lambda_i|$. 

The most important observation that emerges from this part of the study is that \emph{the 2HDM can be valid all the way up to the GUT scale or even the Planck scale without the intervention of any new physics. This is true even if the top quark mass is at the upper end of the currently allowed range.} The additional quartic couplings can counterbalance the effect of the Yukawa coupling threatening vacuum stability, while
still remaining acceptable from the standpoint of perturbativity. It is seen that
we get allowed parameter space for $\L_{UV} = 10^{19}$ GeV corresponding to several values of $\tan\beta$ and $m_{12}$. There is, however, a noticeable correlation - large $m_{12}$ tends to favour small values of $\tan\beta$. For too large an $m_{12}$, the contribution of the extra scalars decouples from the theory. In that case, the RG running of the couplings below that $m_{12}$ is governed by the SM beta functions. In that case, the stability of the electroweak vacuum is again more sensitive to the value chosen for $M_t$.
This has been explicitly checked, for example with $m_{12} = 10^5$ GeV.

The strong coupling constant affects our analysis by determining the initial condition for $g_3$. Current measurements yield a value $0.1184 \pm 0.0007$ for $\alpha_s(M_Z)$. In our
analysis, we have used $\alpha_s(M_Z) = 0.1184$ throughout. However, we demonstrate the
effect of a $3 \sigma$ variation of $\alpha_s(M_Z)$ on the running of $\l_2$, the 
quartic coupling where the effect is expected to be more pronounced compared to the
other ones.

\begin{figure}[!htbp]
\begin{center}
\includegraphics[scale=0.38]{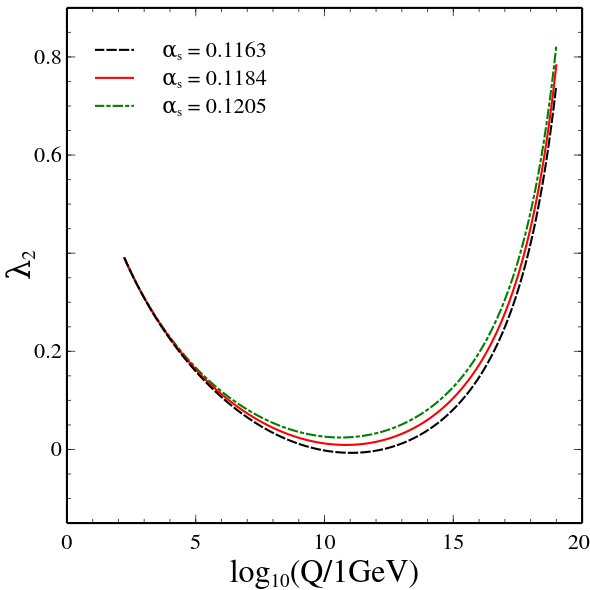}~~~~
\end{center}
\caption{Running of $\l_2$ for three different values for $\alpha_s(M_Z)$ in the soft $\mathbb{Z}_2$ breaking case.}
\label{f:alphas}
\end{figure}

We took $\l_2(M_t) = 0.39$ in Fig.~\ref{f:alphas}. It is seen that the RG running is not significantly altered even by a $3 \sigma$ variation of $\alpha_s(M_Z)$. Hence, for any value of $\alpha_s(M_Z)$ within this band, the parameter spaces will not change in a major fashion, and whatever constraints apply to $\l_2(M_t)$ will continue to be valid rather insensitively to $\alpha_s(M_Z)$.

The implication of having a complex $m_{12}$
in the scalar potential\cite{Grzadkowski:2013rza,Basso:2013wna}
is also investigated here. We rewrite the quadratic part of the scalar potential as,
\be
V_{quad}(\Phi_1,\Phi_2) =
m^2_{11} \Phi_1^{\dagger} \Phi_1
+ m^2_{22} \Phi_2^{\dagger} \Phi_2 -
 |m^2_{12}| \left(e^{i \delta} \Phi_1^{\dagger} \Phi_2 + e^{-i \delta} \Phi_2^{\dagger} \Phi_1\right).
\ee 

The quartic couplings are kept real as in the previous case. The presence of an arbitrary phase $\delta$ in $m_{12}^2$, leads to a charged scalar $H^{+}$, three neutral scalars $H_1$, $H_2$ and $H_3$ which are not eigenstates of CP, and of course the charged
and neutral Goldstone bosons. 
The masses of the neutral scalars can not be obtained in closed form in this case, rather, the corresponding mass matrix has to be diagonalised numerically. In the process of doing that, we choose the lightest neutral scalar, say $H_3$ to be around 125 GeV and the charged scalar to have a mass higher than 315 GeV. The quartic couplings satisfying these conditions are selected and are further constrained by the imposition of the theoretical constraints under RG.

\begin{figure}[!htbp]
\begin{center}
\includegraphics[scale=0.40]{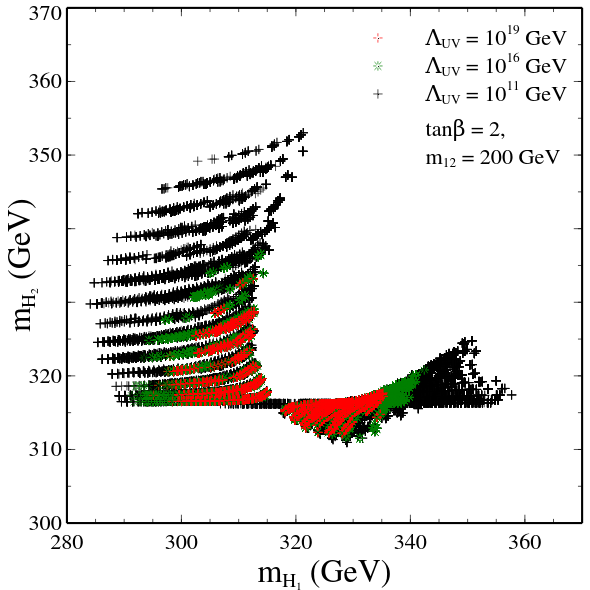}~~~~
\includegraphics[scale=0.40]{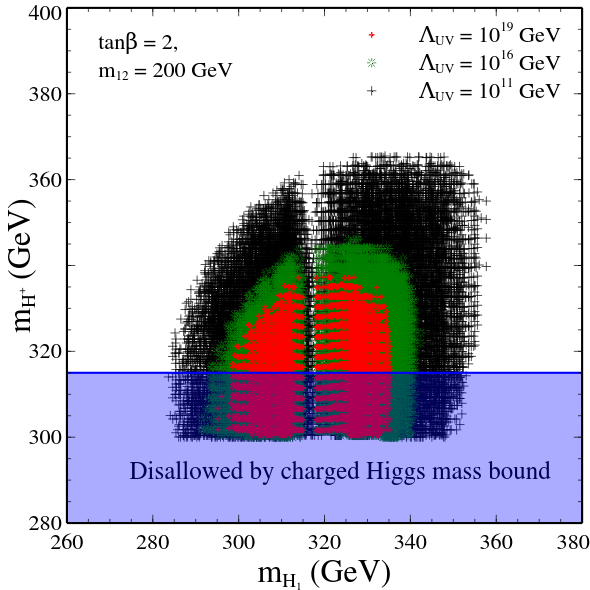}~~~~ \\
\vspace*{0.5cm}
\includegraphics[scale=0.40]{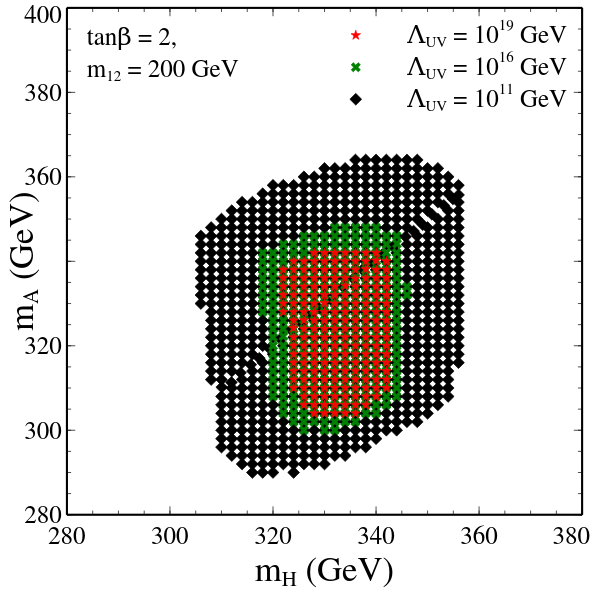}~~~~
\includegraphics[scale=0.40]{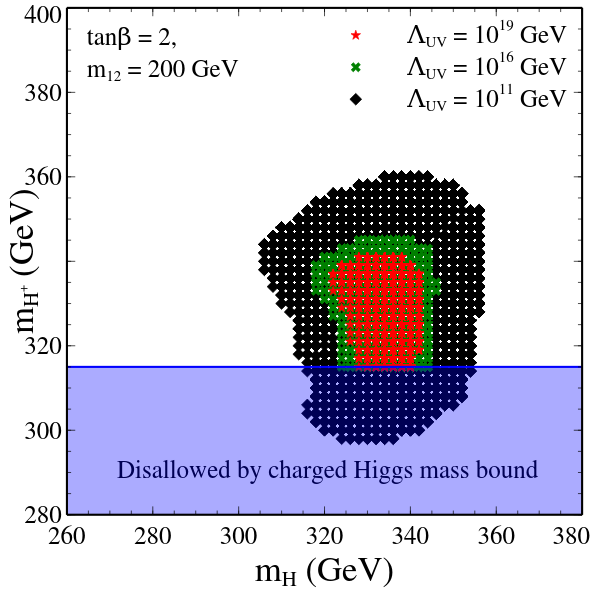}~~~~ \\

\end{center}
\caption{The allowed regions in mass plane as a function of $\L_{UV}$ in the soft $\mathbb{Z}_2$ breaking case.
 The upper and lower two plots correspond to $\delta = \frac{\pi}{4}$ 
 and $\delta = 0$ respectively.}
\label{f:massbounds}
\end{figure}

%
We have chosen the values
$\delta = \frac{\pi}{4}$, $|m_{12}| = 200$ GeV and $\tan\beta$ = 2 as benchmark. 
This choice is illustrative. Constraints on the phase from, say, the electron dipole moment requires full evaluation at each point in the parameter space. For more discussions, we refer the reader to~\cite{Basso:2012st}.
Scatter plots in mass planes are presented in Fig.~\ref{f:massbounds}. For higher $\L_{UV}$, the bounds on the scalar masses become tighter. To make the effect of the added phase in changing the scalar masses, we also show the mass bounds
in the situation with $\delta = 0$.
We would like to emphasize that it is not our purpose here to scan the allowed range of $\delta$ for different values of the mass parameters and quartic couplings. The point that we make is that the validity of this 2HDM up to high scales holds even with a CP-violating phase in the potential. $\delta = \frac{\pi}{4}$ is chosen  as a benchmark for this demonstration. A detailed study of the $\delta$ dependence of the allowed parameter space and its phenomenological implications is the subject of a separate project.  

Our observation therefore is that the regions in the parameter space of a 2HDM, consistent with UV completion at the GUT/Planck scale, are dependent on the phase of the complex parameter(s) of the scalar potential. Together with the less crucial role played by the top mass uncertainty, this is the other important lesson to take home from this section.    

\section {Results with quartic terms breaking the discrete symmetry}
\begin{figure}[!htbp]
\begin{center}
\includegraphics[scale=0.3]{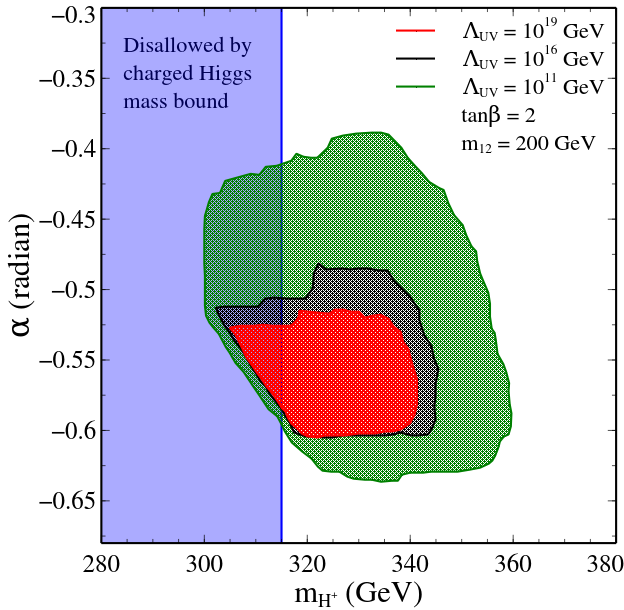}~~~~~~~~
\includegraphics[scale=0.3]{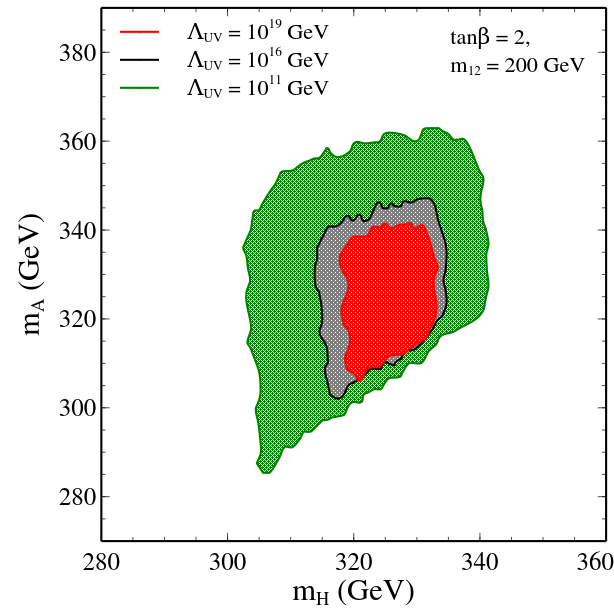}\\
\vspace*{0.5cm}
\includegraphics[scale=0.3]{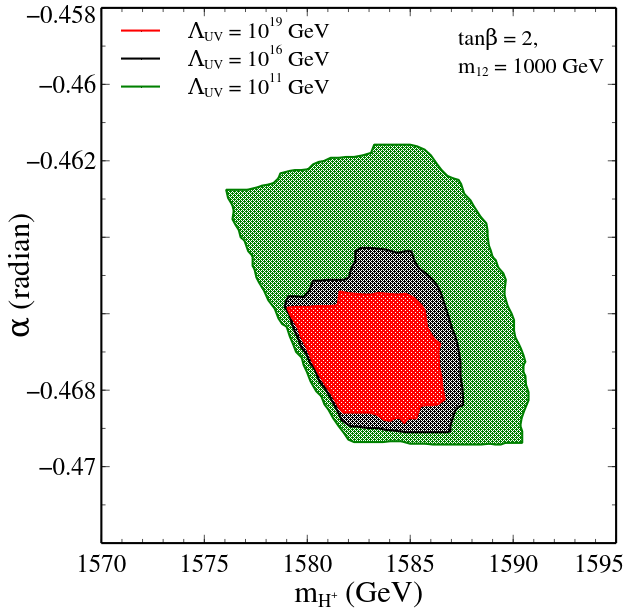}~~~~~~~~
\includegraphics[scale=0.3]{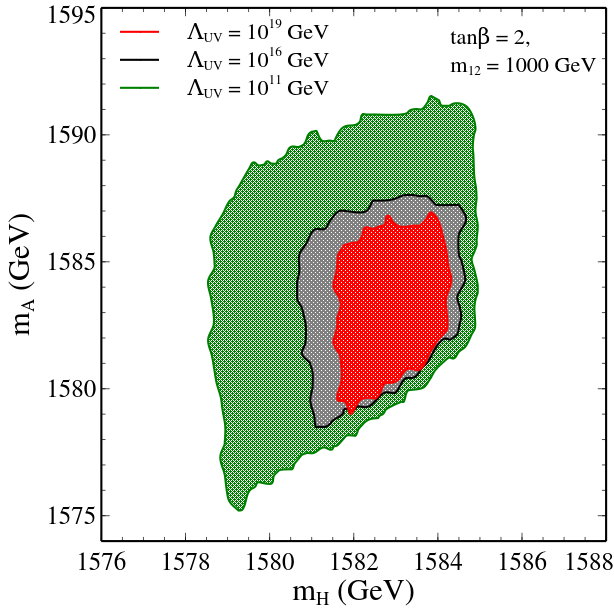}\\
\vspace*{0.5cm}
\includegraphics[scale=0.3]{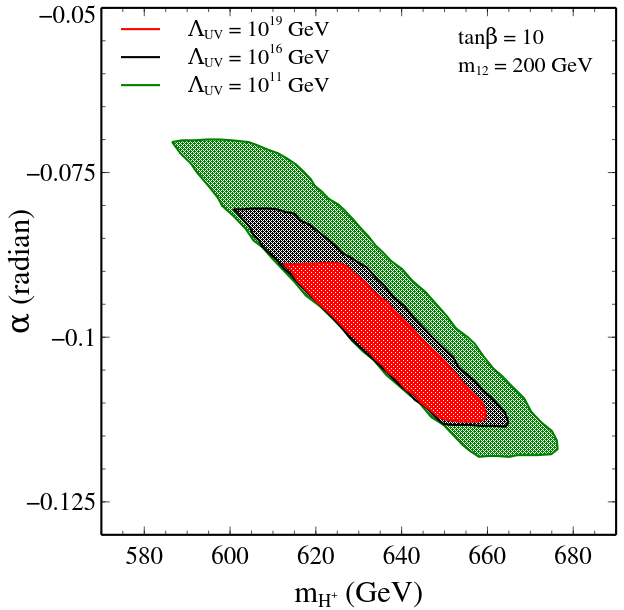}~~~~~~~~
\includegraphics[scale=0.3]{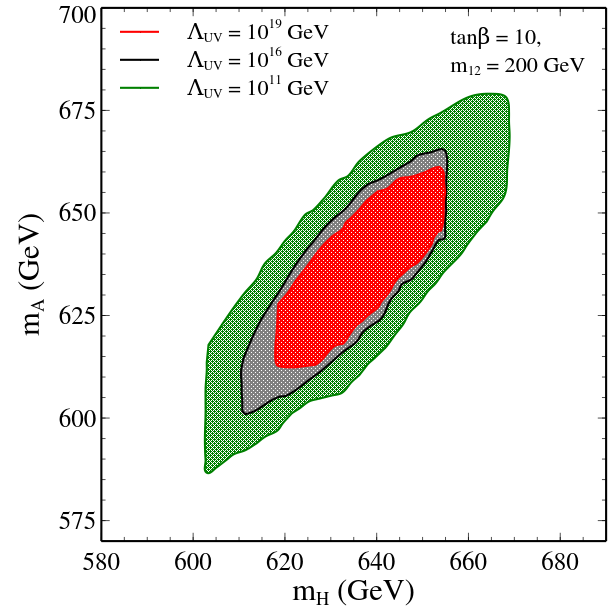}\\
\vspace*{0.5cm}
\includegraphics[scale=0.32]{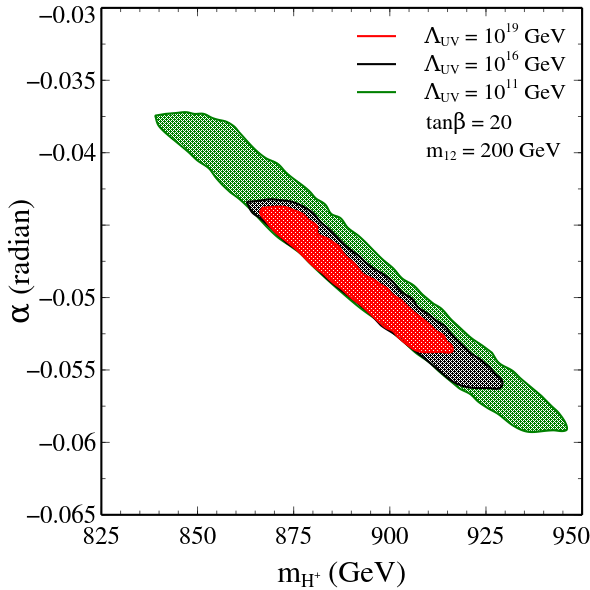}~~~~~~~~
\includegraphics[scale=0.32]{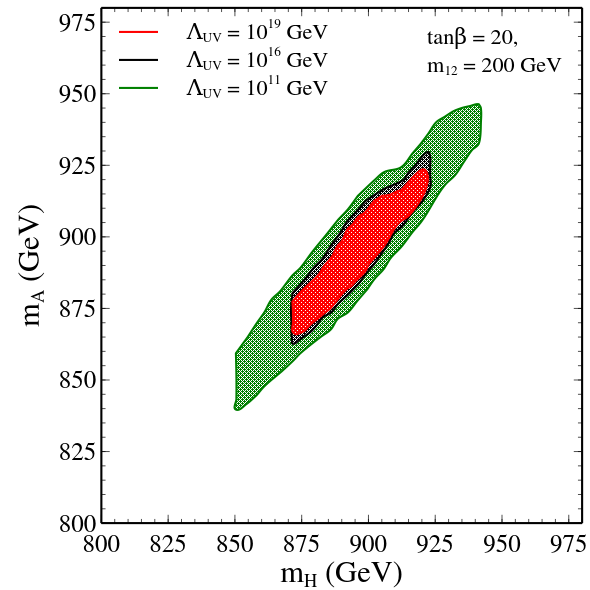}\\
\end{center}
\caption{The allowed parameter spaces for $\Lambda_{UV} = 10^{11}$ (green), $10^{16}$ (grey) and $10^{19}$ GeV (red), in the $\l_6, \l_7 \neq 0$ case. The $\tan\beta$ and $m_{12}$ values are shown in the plots.}
\label{f:w67E111619}
\end{figure}

We now come to the last part of our study where the $\mathbb{Z}_2$ symmetry is broken both at the \emph{soft} and \emph{hard} level (i.e., $m_{12}, \lambda_6 ,\lambda_7 \neq 0$ ). In this case however, the RG running of the various couplings in the theory is different with respect to the \emph{soft} breaking case, owing to the introduction of $\lambda_6$ and $\lambda_7$ (see Appendix \ref{ss:RGE} for the complete set of RG equations). While scanning the $\lambda_i$ parameter space, we try to reduce the number of free parameters
so that the analysis does not become unwieldy. We therefore fix some parameters studied earlier within their allowed ranges. 
In this spirit, we choose $\lambda_1(M_t) = 0.02$ and $\lambda_6(M_t) = \lambda_7(M_t)$ for computational convenience. $\lambda_1(M_t)$ has 
been deliberately chosen to be small 
so that it respects perturbative unitarity even up to the Planck scale. As in the plots shown previously, we present the allowed regions in the $m_H-m_A$ and $m_{H^\pm}-\alpha$ planes which satisfy  all the conditions up to $\Lambda_{UV} = 10^3, 10^{11}, 10^{16}, 10^{19}$ GeV. The benchmarks are $\tan\beta = 2,10,20$ and $m_{12} = 200,1000$ GeV. The results of the scans are shown in Fig.~\ref{f:w67E111619}.

The range over which the scalar masses are distributed can be seen in Fig.~\ref{f:w67E111619}. We note that for $m_{12} = 1$ TeV, the resulting scalar spectrum is almost degenerate. This is precisely due to the fact that the theoretical constraints pin down the allowed values of $\l_i$ to a rather constricted range which also constrains the scalar masses and also the mixing angle in turn. Similarly, for $m_{12} = 200$ GeV, the variation allowed in $\l_i$ causes a variation of $\sim 100$ GeV in the mass of a scalar.    
\begin{figure}[!htbp]
\begin{center}
\includegraphics[scale=0.40]{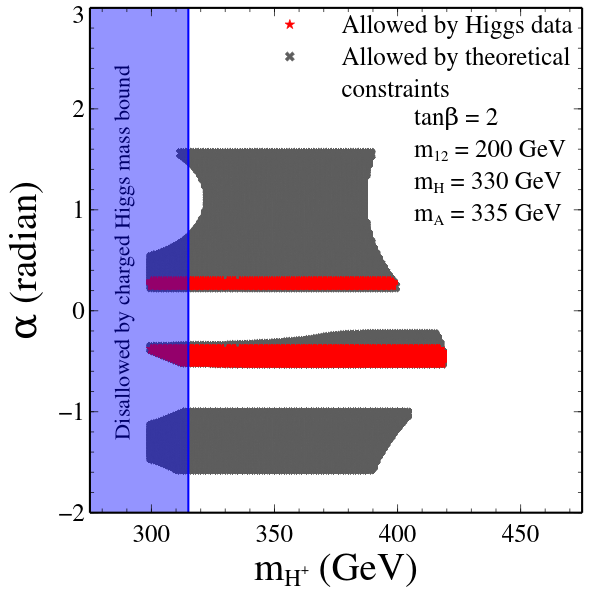}~~~~
\includegraphics[scale=0.40]{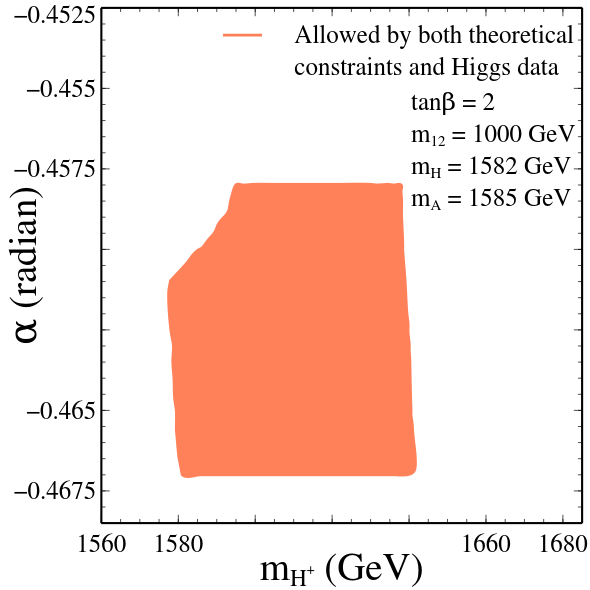}\\
\vspace*{0.5cm}
\includegraphics[scale=0.4]{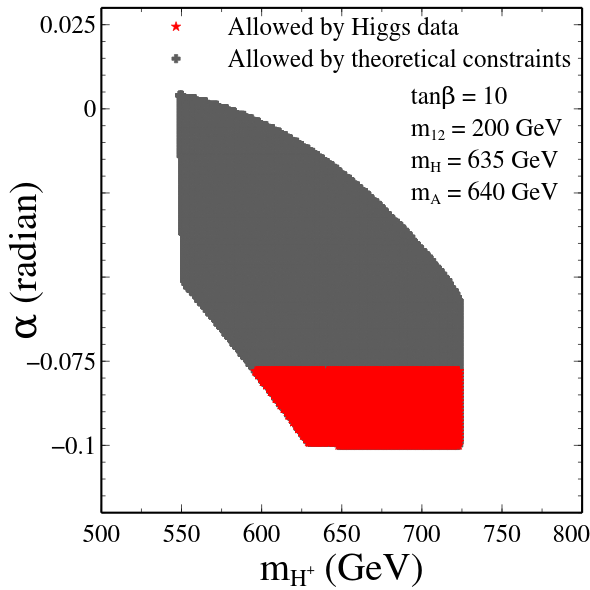}~~~~
\includegraphics[scale=0.4]{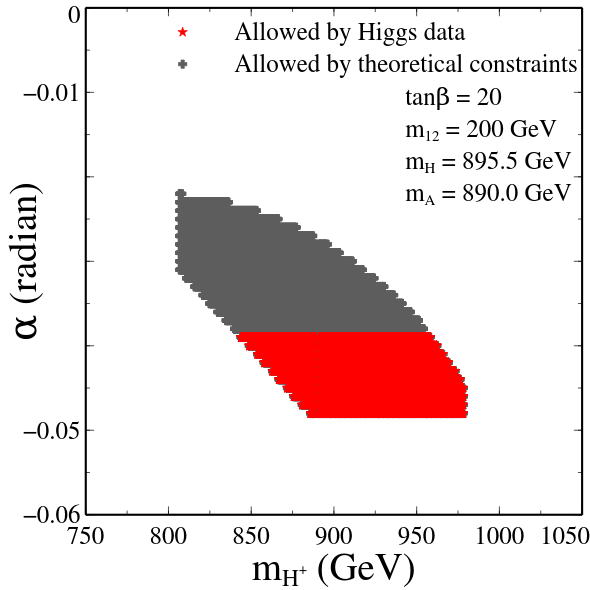}\\
\vspace*{0.5cm}
\end{center}
\caption{Results for $\L_{UV} = 1$ TeV, in the $\l_6, \l_7 \neq 0$ case. The regions in red denote the part of the parameter space allowed by the Higgs data.}
\label{f:67_coll}
\end{figure}

In the case where $\Lambda_{UV} = 10^3$ GeV, we show the subregions in the parameter spaces which are also allowed by the recent Higgs data. Similar to the previous sections, the results have been given in terms of allowed regions in the $m_{H^{\pm}}-\alpha$ plane for specific benchmark values of $m_H$ and $m_A$. The major constraint, however, comes from the signal strength corresponding to $h \rightarrow \gamma \gamma$. It is clearly seen in Fig.~\ref{f:67_coll} that $m_{12} = 1000$ GeV allows for a bigger region in the parameter space that is allowed by the Higgs data at $1\sigma$ level, compared to what $m_{12} = 200$ GeV does. This is obviously expected, given the fact that a high value of $m_{12}$ takes the theory towards the decoupling limit, and thus the 125 GeV Higgs becomes SM-like. Hence, the bounds predicted on the scalar masses and the mixing angle together by the theoretical and collider constraints could be well tested in the next run of the LHC.     

We demonstrate the UV completion of the \emph{hard} $\mathbb{Z}_2$ violating case by showing the RG evolution of the various quartic couplings and stability conditions up to $\Lambda_{UV} = 10^{19}$ GeV. We choose the following initial conditions for the quartic couplings at $\tan\beta = 2$ and $m_{12} = 1000$ GeV,
\bea
&\l_1(M_t) = 0.02,~\l_2(M_t) = 0.48,~\l_3(M_t) = 0.40,~\l_4(M_t) = -0.30,\nonumber \\
&\l_5(M_t) = -0.01,~\l_6(M_t) = -0.05~\mbox{and}~\l_7(M_t) = -0.05~.
\label{e:bc3}
\eea

\begin{figure}[!htbp]
\begin{center}
\includegraphics[scale=0.45]{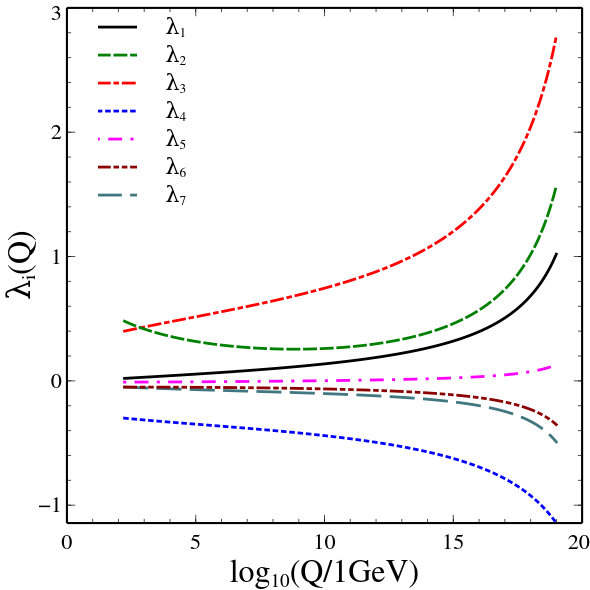}~~~~~~~~
\includegraphics[scale=0.45]{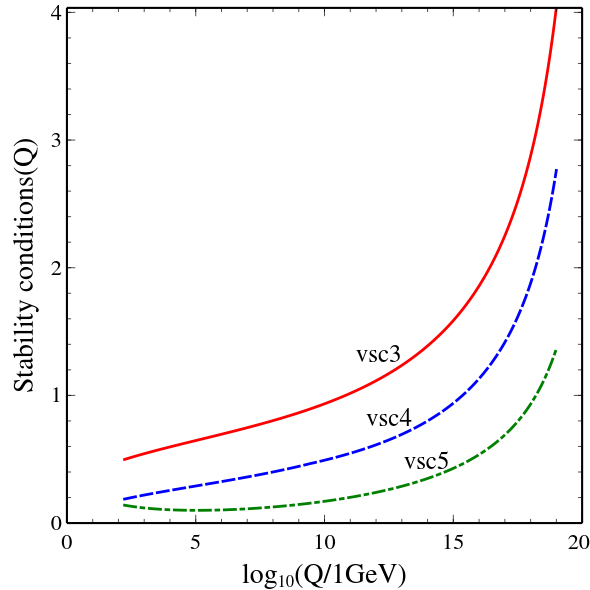}
\end{center}
\caption{RG running of $\lambda_i$ and the stability conditions with the energy scale for $\tan\beta = 2$ and $m_{12} = 1000$ GeV, in the $\l_6, \l_7 \neq 0$ case.}
\label{f:67running}
\end{figure}

These particular initial conditions correspond to, 
\begin{gather}
m_h = 124.62{~\rm GeV}, ~m_H = 1583.33{~\rm GeV}, ~m_{H^{\pm}} = 1585.30{~\rm GeV},\nonumber \\
m_A = 1582.52{~\rm GeV},~\alpha = -0.467{~\rm radian}~,
\end{gather}
and is an allowed point in the parameter space corresponding to the benchmark
$m_{12} = 1000 $ GeV, $\tan\beta = 2$ and $\L_{UV} = 10^{19}$ GeV.
As explained just after Eq.~\ref{e:bc1}, the low-energy boundary values in are just illustrative. In all our quantitative scans (shown in Figs.~\ref{f:diffmt_E3},~\ref{f:mhplal_BP},~\ref{f:wo67E111619},
~\ref{f:diffmt_E19},~\ref{f:wo67E3coll},
~\ref{f:massbounds},~\ref{f:w67E111619} and \ref{f:67_coll}) to determine high-scale validity, a wide range of such boundary conditions are employed. Thus there is nothing fine-tuned about Eqs.~\ref{e:bc1},~\ref{e:bc2} and ~\ref{e:bc3}.  
As shown in Fig.~\ref{f:67running}, $\lambda_3$ increases most sharply whereas $\lambda_2$ first plunges down due to the effect of the $\mathcal{O}(y_{t}^{4})$ term in the RG equation (see Eq.~\ref{e:lam2RG}) and then starts increasing. 
Choosing same initial conditions for $\lambda_6$ and $\lambda_7$ causes their evolutions to become fairly similar. In this section, it should be noted that
the allowed parameter spaces found are not
expected to be exhaustive as we have not scanned over all $\l_i(M_t)$ independently, rather, have put $\l_1(M_t) = 0.02$ and $\l_6(M_t) = \l_7(M_t)$ while doing so. However,
given the similar structure of the 1-loop beta functions of $\l_6$ and $\l_7$ (see Eq.~\ref{eq:dl6} and~\ref{eq:dl7}), the bounds
obtained on them would have not substantially changed even if an independent scanning would have been done.  

\section{Summary and Conclusions} 

We set out to investigate the high-scale behaviour of a 2HDM. The results are
illustrated in the context of a Type-II scenario. We have used the theoretical constraints of perturbativity, unitarity and vacuum stability to constrain the parameter space of the model. The relatively less stringent constraints from oblique
parameters, and also the LHC constraints on the signal strength in each decay channel
of a Higgs around 125 GeV have also been taken into account. 

We find that a 2HDM with a discrete $\mathbb{Z}_2$ symmetry (thereby  
forbidding some cross-terms in the two doublets in the potential) cannot
be valid beyond 10 TeV, since otherwise the requirement of keeping one neutral 
scalar mass around 125 GeV cannot be met. With the discrete symmetry
broken, on the other hand, it is possible to fulfill all the constraints
over a  much larger region of the parameter space. Thus the theory with a 2HDM
can distinctly be valid up to energies as high as $10^{16}$ GeV or even the Planck scale,
without the intervention of any additional physics. This feature
holds irrespectively of the uncertainty in the measured value of the top quark mass,
which is in contrast to what is expected in the standard model with a single Higgs
doublet. In addition, high-scale validity
 of this scenario is not affected by the uncertainty  
 in the strong coupling $\alpha_s (M_Z)$. The effect of a CP-violating phase in the potential is also
considered, it is found that one can find regions in the parameter space valid up to high scales for at least one illustrative value (viz. $\frac{\pi}{4}$) of the phase. The allowed regions of the parameter space, in
terms of the various quartic couplings as well as the scalar mass eigenvalues
are presented by us in detail, in the light of theoretical as well
as collider bounds. The inclusion of $\mathbb{Z}_2$-breaking quartic couplings, too,
is found to retain the high-scale validity of the theory over a large region.

Though the study is based on a type II 2HDM, many of the results obtained here are expected to hold for a more general 2HDM as well. A situation where some departure can take place is, for example one where the Yukawa coupling of the bottom quark becomes comparable to, or more than, that of the top quark. One possibility to explore in such a case is to look for those regions where the large number of quartic couplings can rescue the scenario from an unstable vacuum. The results presented here are based on one-loop RG equations, in consonance with most similar studies in the context of 2HDM.


It should also be noted that we call those regions in the parameter space
as \emph{allowed}, where the vacuum is strictly \emph{stable}. The inclusion of a metastable vacuum, with lifetime greater than the age of the universe, will lead to
larger allowed regions.  

On the whole, our conclusion is that it is possible to validate a 2HDM till scales as high as the Planck mass without any additional physics.
While the issue of naturalness remains unaddressed in this statement, 
it is interesting to see that no current experimental measurement or theoretical restriction can affect high-scale validity, which is not the case for the single-doublet scenario.
   
\section*{Acknowledgements\,:}
We thank Shankha Banerjee, Joydeep Chakrabortty, Samrat Kadge and Sourov Roy for many helpful discussions. This work was partially supported by the funding available from the Department of Atomic Energy, Government of India, for the Regional Centre for Accelerator-based Particle Physics, Harish-Chandra Research Institute. Computational work for this study was partially carried out at the cluster computing facility in the Harish-Chandra Research Institute (http://cluster.hri.res.in).

\section*{Appendix}
\appendix
\label{s:appen}
\section{Renormalization group (RG) equations}
\label{ss:RGE}
The RG equations for the gauge couplings, for this model, are given by \cite{Branco:2011iw},
\besub
\bea
16\pi^2 \frac{dg_s}{dt} &=& - 7 g_s^3,
\\
16\pi^2 \frac{dg}{dt} &=& - 3 g^3,
\\
16\pi^2 \frac{dg^{\prime}}{dt} &=& 7 {g^\prime}^3.
\eea
\eesub

Since we want to avoid CP violation coming from the quartic sector of the Higgs potential, we choose to keep $ \lambda_{i} ~(i=1,\ldots,7)$ real. In that case, the quartic couplings evolve according to, 
\besub
\bea
\label{e:lam1RG}
16\pi^2 \frac{d\lambda_1}{dt} &=&
12 \lambda_1^2 + 4 \lambda_3^2 + 4 \lambda_3 \lambda_4 + 2 \lambda_4^2
+ 2  \lambda_5^2 + 24 \lambda_6^2 + \frac{3}{4}(3g^4 + g^{\prime 4} +2 g^2 g^{\prime 2})\nonumber \\
 & & -\lambda_1 (9 g^2 + 3 g^{\prime
2} - 12 y_b^2 - 4 y_{\tau}^2 ) - 12 y_b^4 - 4 y_{\tau}^4\,, \\
\label{e:lam2RG}
16\pi^2 \frac{d\lambda_2}{dt} &=&
12 \lambda_2^2 + 4 \lambda_3^2 + 4 \lambda_3 \lambda_4 + 2 \lambda_4^2
+ 2 \lambda_5^2 + 24 \lambda_7^2 \nonumber \\
 & &
+\
\frac{3}{4}(3g^4 + g^{\prime 4} +2g^2 g^{\prime 2}) -3\lambda_2
(3g^2 +g^{\prime 2} -  4 y_t^2) - 12 y_t^4\,, \\
16\pi^2 \frac{d\lambda_3}{dt}  &=&
\left( \lambda_1 + \lambda_2 \right) \left( 6 \lambda_3 + 2 \lambda_4 \right)
+ 4 \lambda_3^2 + 2 \lambda_4^2
+ 2 \lambda_5^2
+ 4 \left(\lambda_6^2 + \lambda_7^2 \right) \nonumber \\
& & + 16\left( \lambda_6 \lambda_7 \right)
+\frac{3}{4}(3g^4 + g^{\prime 4} -2g^2 g^{\prime 2}) \nonumber \\
& & - \lambda_3
(9g^2 + 3g^{\prime 2}- 6 y_t^2 - 6 y_b^2 - 2 y_{\tau}^2) - 12 y_t^2 y_b^2\,, \\
16\pi^2 \frac{d\lambda_4}{dt}  &=&
2 \left( \lambda_1 + \lambda_2 \right) \lambda_4
+ 8 \lambda_3 \lambda_4 + 4 \lambda_4^2
+ 8 \lambda_5^2
+ 10 \left( \lambda_6^2 + \lambda_7^2 \right)
+ 4\left( \lambda_6 \lambda_7\right) \nonumber \\
& &
+\ 3g^2 g^{\prime 2} - \lambda_4 (9g^2 + 3g^{\prime
2}- 6 y_t^2 - 6 y_b^2 - 2 y_{\tau}^2) + 12 y_t^2 y_b^2\,,
\label{eq:dl4} \\
16\pi^2 \frac{d\lambda_5}{dt}  &=&
\left( 2 \lambda_1 + 2 \lambda_2 + 8 \lambda_3 + 12 \lambda_4 \right) \lambda_5
+ 10 \left( \lambda_6^2 + \lambda_7^2 \right) + 4 \lambda_6 \lambda_7
\nonumber \\
 & &
- \ \lambda_5 (9g^2 + 3g^{\prime 2}- 6 y_t^2 - 6 y_b^2 - 2 y_{\tau}^2)\,, \label{eq:dl5} \\
16\pi^2 \frac{d\lambda_6}{dt}  &=&
\left( 12 \lambda_1 + 6 \lambda_3 + 8 \lambda_4 \right) \lambda_6
+ \left( 6 \lambda_3 + 4 \lambda_4 \right) \lambda_7
+ 10 \lambda_5 \lambda_6 + 2 \lambda_5 \lambda_7\nonumber \\
 & &
-\ \lambda_6 (9g^2 + 3g^{\prime 2}- 9 y_b^2 - 3 y_t^2 - 3 y_{\tau}^2)\,, 
\label{eq:dl6} \\
16\pi^2 \frac{d\lambda_7}{dt}  &=&
\left( 12 \lambda_2 + 6 \lambda_3 + 8 \lambda_4 \right) \lambda_7
+ \left( 6 \lambda_3 + 4 \lambda_4 \right) \lambda_6
+ 10 \lambda_5 \lambda_7 + 2 \lambda_5 \lambda_6\nonumber \\
 & &
-\ \lambda_7 (9g^2 + 3g^{\prime 2} - 9 y_t^2 - 3 y_b^2 - y_{\tau}^2)\,.
\label{eq:dl7} 
\eea
\eesub
For the Yukawa couplings the corresponding set of RG equations are,
\besub
\bea
 16\pi^2 \frac{dy_b}{dt} &=& y_{b}\left(-8g_s^2 - \frac94 g^2 - \frac{5}{12} g^{\prime 2}+
 \frac92 y_{b}^2 +y_{\tau}^2 + \frac12 y_{t}^2\right)\,,\\
 16\pi^2 \frac{dy_t}{dt} &=& y_{t}\left(-8g_s^2 - \frac94 g^2 - \frac{17}{12} g^{\prime 2}+
 \frac92 y_{t}^2 + \frac12 y_{b}^2\right)\,,\\
 16\pi^2 \frac{dy_{\tau}}{dt} &=& y_{\tau}\left(-\frac94 g^2 - \frac{15}{4} g^{\prime 2}+
 3 y_{b}^2 + \frac52 y_{\tau}^2\right)\,.
\eea
\eesub

\section{Unitarity bounds}
\label{ss:LQT}
We perform a coupled channel analysis of $2 \rightarrow 2$ scattering involving
fields in the scalar sector, to the leading order.
The basis of neutral two-particle states is given by,
\be
\left\lbrace w_{1}^{+}w_{2}^{-}, w_{2}^{+}w_{1}^{-}, h_{1}z_{2}, h_{2}z_{1}, z_{1}z_{2}, h_{1}h_{2}, h_{1}z_{1}, h_{2}z_{2}, w_{1}^{+}w_{1}^{-}, w_{2}^{+}w_{2}^{-}, \frac{z_{1}z_{1}}{\sqrt{2}}, \frac{z_{2}z_{2}}{\sqrt{2}}, \frac{h_{1}h_{1}}{\sqrt{2}}, \frac{h_{2}h_{2}}{\sqrt{2}}\right\rbrace
\ee
For the general $\lambda_6, \lambda_7 \neq 0$ case, the $(14\times 14)$ two-particle scattering matrix is given as follows:
\bea
\mathcal{M}_{NC}= \begin{pmatrix}
\mathcal{A}_{7\times 7} & \mathcal{B}_{7\times 7} \\ \mathcal{B}^{\dagger}_{7\times 7} & \mathcal{C}_{7\times 7}
\end{pmatrix} \,,
\eea
where $\mathcal{A}, \mathcal{B}$ and $\mathcal{C}$ are given by,
\bea
\mathcal{A}_{7\times 7} = \begin{footnotesize} \begin{pmatrix}
 \l_3+\l_4 & 2\l_5 & \frac{i}{2}(\l_4-\l_5) & \frac{i}{2}(-\l_4+\l_5) & \frac12(\l_4+\l_5) & \frac12(\l_4+\l_5) & 0 \\ 
 2\l_5 & \l_3+\l_4 & \frac{i}{2}(-\l_4+\l_5) & \frac{i}{2}(\l_4-\l_5) & \frac12(\l_4+\l_5) & \frac12(\l_4+\l_5) & 0 \\
\frac{i}{2}(-\l_4+\l_5) & \frac{i}{2}(\l_4-\l_5) & (\l_3+\l_4-\l_5) & \l_5 & 0 & 0 & \l_6 \\
\frac{i}{2}(\l_4-\l_5) & \frac{i}{2}(-\l_4+\l_5) & \l_5 & (\l_3+\l_4-\l_5) & 0 & 0 & \l_6 \\
\frac12(\l_4+\l_5) & \frac12(\l_4+\l_5) & 0 & 0 & (\l_3+\l_4+\l_5) & \l_5 & 0 \\
\frac12(\l_4+\l_5) & \frac12(\l_4+\l_5) & 0 & 0 & \l_5 & (\l_3+\l_4+\l_5) & 0 \\
0 & 0 & \l_6 & \l_6 & 0 & 0 & \l_1
\end{pmatrix},
\end{footnotesize} \nonumber
\eea
\bea
\mathcal{B}_{7\times 7} =  \begin{small} \begin{pmatrix}
0 & 2\l_6 & 2\l_7 & \frac{\l_6}{\sqrt{2}} & \frac{\l_7}{\sqrt{2}} & \frac{\l_6}{\sqrt{2}} & \frac{\l_7}{\sqrt{2}} \\
0 & 2\l_6 & 2\l_7 & \frac{\l_6}{\sqrt{2}} & \frac{\l_7}{\sqrt{2}} & \frac{\l_6}{\sqrt{2}} & \frac{\l_7}{\sqrt{2}} \\
\l_7 & 0 & 0 & 0 & 0 & 0 & 0 \\
\l_7 & 0 & 0 & 0 & 0 & 0 & 0 \\
0 & \l_6 & \l_7 & \frac{3\l_6}{\sqrt{2}} & \frac{3\l_7}{\sqrt{2}} & \frac{\l_6}{\sqrt{2}} & \frac{\l_7}{\sqrt{2}} \\
0 & \l_6 & \l_7 & \frac{\l_6}{\sqrt{2}} & \frac{\l_7}{\sqrt{2}} & \frac{3\l_6}{\sqrt{2}} & \frac{3\l_7}{\sqrt{2}} \\
\l_5 & 0 & 0 & 0 & 0 & 0 & 0
\end{pmatrix},
\end{small}  \nonumber
\eea
\bea
\mathcal{C}_{7\times 7} = \begin{footnotesize} \begin{pmatrix}
 \l_2 & 0 & 0 & 0 & 0 & 0 & 0\\ 
 0 & 2\l_1 & (\l_3+\l_4) & \frac{\l_1}{\sqrt{2}} &  \frac{\l_3}{\sqrt{2}} &  \frac{\l_1}{\sqrt{2}} &  \frac{\l_1}{\sqrt{2}}\\
 0 & (\l_3+\l_4) & 2\l_2 & \frac{\l_3}{\sqrt{2}} &  \frac{\l_2}{\sqrt{2}} &  \frac{\l_3}{\sqrt{2}} &  \frac{\l_2}{\sqrt{2}}\\
0 & \frac{\l_1}{\sqrt{2}} & \frac{\l_3}{\sqrt{2}} & \frac{3\l_1}{2} & \frac{1}{2}(\l_3+\l_4+\l_5) & \frac{\l_1}{2} & \frac{1}{2}(\l_3+\l_4-\l_5) \\
0 & \frac{\l_3}{\sqrt{2}} & \frac{\l_2}{\sqrt{2}} & \frac{1}{2}(\l_3+\l_4+\l_5) & \frac{3\l_2}{2} & \frac{1}{2}(\l_3+\l_4-\l_5) & \frac{\l_2}{2} \\
0 & \frac{\l_1}{\sqrt{2}} & \frac{\l_3}{\sqrt{2}} & \frac{\l_1}{2} & \frac{1}{2}(\l_3+\l_4-\l_5) & \frac{3\l_1}{2} & \frac{1}{2}(\l_3+\l_4+\l_5) \\
0 & \frac{\l_3}{\sqrt{2}} & \frac{\l_2}{\sqrt{2}} & \frac{1}{2}(\l_3+\l_4-\l_5) & \frac{\l_2}{2} & \frac{1}{2}(\l_3+\l_4+\l_5) & \frac{3\l_2}{2} \\
\end{pmatrix}.
\end{footnotesize}  \nonumber
\eea
The constraint imposed by unitarity is then given by $|a_{i}| \leq 8 \pi$, where $a_i~(i=1,\ldots,14)$ are eigenvalues of the matrix $\mathcal{M}$. The eigenvalues of $\mathcal{M}$ are evaluated numerically in the present study. However, in the absence of \emph{hard} $\mathbb{Z}_2$ breaking, i.e., when $\lambda_6,\lambda_7 = 0$, the matrix decomposes into blocks and analytical expressions for its eigenvalues can be obtained in simple forms which are listed below. 
\besub
\bea
a_{\pm}&=& \frac32(\lambda_1+\lambda_2)\pm \sqrt{\frac94 (\lambda_1-\lambda_2)^2+(2\lambda_3+\lambda_4)^2},\\
b_{\pm}&=& \frac12(\lambda_1+\lambda_2)\pm \sqrt{\frac14 (\lambda_1-\lambda_2)^2+\lambda_4^2},\\
c_{\pm}&=& d_{\pm} = \frac12(\lambda_1+\lambda_2)\pm \sqrt{\frac14 (\lambda_1-\lambda_2)^2+\lambda_5^2},\\
e_1&=&(\lambda_3 +2\lambda_4 -3\lambda_5),\\
e_2&=&(\lambda_3 -\lambda_5),\\
f_1&=& f_2 = (\lambda_3 +\lambda_4),\\
f_{+}&=& (\lambda_3 +2\lambda_4 +3\lambda_5),\\
f_{-}&=& (\lambda_3 +\lambda_5).
\eea
\label{e:LQTeval}
\eesub
The matrix corresponding to the overall singly charged states,
\be
 \{ h_1 w_1^{+}, h_2 w_1^{+}, z_1 w_1^{+}, z_2 w_1^{+}, h_1 w_2^{+}, h_2 w_2^{+}, z_1 w_2^{+}, z_2 w_2^{+}\}
\ee 
  is given by,
\bea
\mathcal{M}_{CC} =  \begin{pmatrix}
 \l_1 & \l_6 & 0 & 0 & \l_6 & \frac{1}{2}(\l_4+\l_5) & 0 & \frac{i(\l_4-\l_5)}{2}\\ 
 \l_6 & \l_3 & 0 & 0 & \frac{\l_4+\l_5}{2} & \l_7 & \frac{-i(\l_4-\l_5)}{2} & 0\\
 0 & 0 & \l_1 & \l_6 & 0 & \frac{-i(\l_4-\l_5)}{2} & \l_6 & \frac{\l_4+\l_5}{2}\\
0 & 0 & \l_6 & \l_3 & \frac{i(\l_4-\l_5)}{2} & 0 & \frac{\l_4+\l_5}{2} & \l_7 \\
\l_6 & \frac{\l_4+\l_5}{2} & 0 & \frac{-i(\l_4-\l_5)}{2} & \l_3 & \l_7 & 0 & 0 \\
\frac{\l_4+\l_5}{2} & \l_7 & \frac{i(\l_4-\l_5)}{2} & 0 & \l_7 & \l_2 & 0 & 0 \\
0 & \frac{i(\l_4-\l_5)}{2} & \l_6 & \frac{\l_4+\l_5}{2} & 0 & 0 & \l_3 & \l_7\\
\frac{-i(\l_4-\l_5)}{2} & 0 & \frac{\l_4+\l_5}{2} & \l_7 & 0 & 0 & \l_7 & \l_2 \\
\end{pmatrix}. \nonumber
\eea
Again for the case $\l_6,\l_7=0$, the eigenvalues of $\mathcal{M}_{CC}$ are, $b_{\pm}, c_{\pm}, e_{2}, f_{1}, f_{-}$ and $p = (\l_3-\l_4)$.   

\bibliographystyle{JHEP}
\bibliography{ref.bib}       
\end{document}